\documentclass[a4,12pt]{article}
\usepackage{xcolor}
\usepackage{graphicx}
\usepackage{dcolumn}
\usepackage{bm}
\usepackage{amsmath,amssymb}
\usepackage[english]{babel}
\usepackage[utf8]{inputenc}
\usepackage{fancyhdr} 
\graphicspath{ {./images/} }
\begin{document}
\begin{center}
{\bf Dedicated to the birth centenary of the Father of Nation, Bangabandhu Sheikh Mujibur Rahman}
\end{center}
%
{\bf {\large MICROSCOPIC ORIGIN OF IMMISCIBILITY AND SEGREGATION IN LIQUID BINARY ALLOYS}} \\\\
{\bf G. M. BHUIYAN} \\\\
{\it Department of Theoretical Physics, University of Dhaka, Bangladesh} \\ 
\vspace{-2mm}\\
{\it \small{$^{*}$Correspondence E-mail: gbhuiyan@du.ac.bd}} \\ 
\vspace{1mm}\\
{\bf ABSTRACT}\\ 
\vspace{-3mm} \\
{\footnotesize Microscopic description in the study of immiscibility and segregating properties of liquid metallic binary alloys 
has gained a renewed scientific and technological interests during the last eight years for the physicists, metallurgists and chemists.
Especially, in understanding the basic mechanisms, from the point of interionic interaction, and how and why segregation in some metallic alloys 
takes place at and under certain thermodynamic state specified by temperature and pressure. An overview of the theoretical  and
experimental works done by different authors or groups in the area of segregation combining electronic theory of metals, statistical mechanics and 
the perturbative approach is presented in this review. Main attention in this review is focused on the static effects such as the effects of energy of mixing, 
enthalpy of mixing, entropy of mixing and understanding the critical behaviour of segregation of alloys from the microscopic theoretical approach. 
Investigation of  segregating properties from the dynamic effects such as from the
effects of shear viscosity and diffusion coefficient is just becoming available. However, we have restricted this review only on 
static effects and their variation of impacts on different alloys.}    
%
%
\newpage
\pagestyle{fancy}
\fancyhead{}
\fancyhead[RO]{G M BHUIYAN} 
\fancyhead[LO]{\thepage}
\section{Introduction}
Some advancement in understanding the segregating properties, miscibility gap, demixing tendency etc. of some metallic binary alloys, has been
made,  
so far, from the empirical models [1-3], and phenomenological theories [4,5] in conjunction with arbitrary concepts of association and dissociation[6-8].
Experimental data for some liquid segregating alloys [9-20] plays the pivotal role to arouse the interest in theoretical study, in particular to understand the critical behaviours. 
This knowledge is required to find the possible application of segregating materials to innovate technology and to industries for car engines, electrical contacts and switches,
separation of impurities from the iron melts, ceramic industries, cosmetic and the food industry.

Known signatures of the existence of immiscibility, segregation, miscibility gap, and critical properties of segregating alloys are 
deviation from Roult's law,  concave downward of  the free energy of mixing profile [21-25], 
concave upward of entropy [24,25] and entalpy of mixing [25], large density fluctuation displayed by concentration-concentration structure factors [26], 
large difference in partial coordination numbers [27]
derived by using partial pair correlation functions $g_{ii}(r)$ and $g_{12}$, sudden sharp bending of the atomic transport properties as a function of concentration [25],
positivity of short range order parameter [28-32], exhibiting some sort of scaling laws [25] etc.
But, understanding of the actual mechanisms involved behind these signatures is a great challenge to physicists, metallurgists and the material engineers.

Segregating properties of liquid binary alloys may be studied microscopically from the static [21-24]and dynamic effects [25]. The static effects may be observed from the thermodynamic
properties of mixing, coordination number derived from structural properties etc. The dynamic effects can be seen from the atomic transport properties such as
coefficient of viscosity and diffusion coefficient, and the electronic transport properties such as electrical resistivity [26]. In this review, a microscopic 
theoretical approach that involves the electronic theory of metals [23,33-35], perturbation theory [36,38], 
the hard sphere reference system [39,40] and the statistical mechanics. Electron ion interaction is described by a local pseudopotential, the interionic 
pair interaction is derived from the energy band structure which is finally employed to evaluate static structure of liquid metals and their alloys.
Specifically, the form factors of the pseudo potential is used to find effective pair potentials and the volume dependent contribution to the free energy.
The knowledge of pair potentials is essential to have pair correlation functions, the energy of the reference system 
\newpage
\pagestyle{fancy}
\fancyhead{}
\fancyhead[LO]{MICROSCOPIC ORIGIN OF IMMISCIBILITY $\cdots$} 
\fancyhead[RO]{\thepage}
\hspace{-6mm}and that of the attractive tail [33,37].

There are many liquid metallic binary alloys which exhibit miscibility gap or segregation at certain thermodynamic state. Some of these alloys are Li-Na,
Al-In, Al-Pb, Al-Bi, Zn-Bi, Bi-Ga, Ga-Pb, Ga-Hg, Pb-Sn, Fe-Cu, Co-Cu, Cu-Pb etc. Of them, only a few systems such as Al-In, Al-Bi, Zn-Bi, Fe-Cu, Co-Cu 
are systematically studied employing microscopic theory, of course, empirical or phenomenological theories are applied to study some other systems[26]. For 
Al-In, Al-Bi, Zn-Bi, Fe-Cu, Co-Cu liquid binary alloys 
break down details are available (see below), from which one can analyze which component of the interionic interaction contributes how much or dominates in making the segregation to happen.
Finally, comparison of these results with the experimental data would help understand the origin of segregation from the microscopic point of view and also the limitations
of the employed theoretical approaches.

Very little efforts have been spent so far in the study of immiscibility or segregating behaviour of liquid metallic alloys from the effects of dynamic properties 
when it is compared with that of the static effects.
 So it demands further to have considerably more studies in this direction. For this purpose the easyest way is to invoke the Rice-Allnatt theory for atomic transport properties.
 Because, analytic expressions
for shear viscosity and diffusion coefficient are already available for elemental [41,42] and binary alloys[43].

The layout of this review is as follows. Relevant theories are briefly discussed in section 2. Section 3 is devoted to the results obtained from the empirical
and phenomenological theories. Results for the segregating properties for different alloys calculated from the microscopic theories are presented in section 4.
A brief comparison of the impacts of the interionic interactions on different alloy systems is done and analysed in the concluding section 5.
\section{Theory}
Different theories relevant to the present review article  are briefly presented below.
\subsection{Thermodynamic relations involved}
A macrostate of a condensed system may be described by four independent variables. These are pressure, $p$, volume, $V$, temperature, $T$, and entropy, $S$.
Here, $p$ and $V$ form a pair representing the mechanical degrees of freedom, 
\newpage
\pagestyle{fancy}
\fancyhead{}
\fancyhead[RO]{G M BHUIYAN} 
\fancyhead[LO]{\thepage}
$\hspace{-6mm}$and $T$ and $S$ form another pair representing the thermal degrees of freedom.
Any two of the four variables may be chosen in six different ways. Of them four pair of variables are (p,T), (p,s), (V,T) and (V,S), each of which contains one variable
from  the mechanical and another from the thermal degrees of freedom. Thermodynamic functions constructed by these pairs are Gibbs free energy, G(p,T), enthalpy, H(p,S), 
Helmholtz free energy, F(V,T), and the internal energy, E(V,S).

The free energy in thermodynamics is the amount of energy of the system free to work. Internal energy of a system is the sum of kinetic energy,
potential energy, rotational energy and the vibrational energy etc. Of course, in the magnetic systems the magnetic energy [44,45] and
for a finite sized sample the surface energy correction to be counted in the above functions [46]. For a monoatomic systems (also in random binary alloys) there are no
rotational and vibrational energy contribution in general. However, the above thermodynamic functions are not independent to each other. They are
rather interconnected. The Helmholtz free energy (for the bulk) is (dropping the variables for brevity)
\begin{eqnarray}
 F = E - TS\,  .
\end{eqnarray}
The Gibbs free energy
\begin{eqnarray}
 G = F + pV = E-TS+p V\,  .
\end{eqnarray}
The enthalpy
\begin{eqnarray}
 H=E + pV \,.
\end{eqnarray}
And the change in the internal energy
\begin{eqnarray}
 dE = T dS - p dV\, .
\end{eqnarray}
From the theoretical point of view, we can further analyze the above relations. for example, for zero pressure, i.e. at $p=0$,
\begin{eqnarray}
 G&=& F  \\
H &=& E \, .
\end{eqnarray}

Again, most of the experimental data for thermodynamical quantities available in the literature are at standard temperature and pressure. 
In one 
\newpage
\pagestyle{fancy}
\fancyhead{}
\fancyhead[LO]{MICROSCOPIC ORIGIN OF IMMISCIBILITY $\cdots$} 
\fancyhead[RO]{\thepage}
$\hspace{-6mm}$atmospheric pressure the value of the product of $pV$ appears to be very small when compared with other terms of the above thermodynamic functions.
So, in one or two atmospheric pressure or less one can write
\begin{eqnarray}
 H \approx E\,; \hspace{1cm} G \approx F \, .
\end{eqnarray}
We note here that other physical quantities such as heat capacity, compressibility etc. can be derived from equations (1) to (3) [47-49].

For binary alloys the free energy of mixing is 
\begin{eqnarray}
\Delta F = F_{alloy} -\sum_{a} C_{a} F_{a} 
\end{eqnarray}
where $F_{alloy}$ is the free energy of the alloy, $C_{a}$ is the concentration of the $a$-th component and, $F_{a}$ is the free energy of the $a$-th element 
in the same thermodynamic state. Similarly the enthalpy of mixing may be expressed as
\begin{eqnarray}
\Delta H = H_{alloy} -\sum_{a} C_{a} H_{a} \,,
\end{eqnarray}
and the entropy of mixing as
\begin{eqnarray}
\Delta S = S_{alloy} -\sum_{a} C_{a} S_{a} \,.
\end{eqnarray}

It is worth noting that, the thermodynamics is a phenomenological subject because all relations in thermodynamics are obtained just looking
at the experimental results. The only way to have microscopic description of the thermodynamic quantities is through
the statistical mechanics [48,49]. 
\section{Microscopic theory for metallic systems}
\subsection{The pair correlation function}
Let us consider $N$ ions each of valence $Z$ are there in a volume $V$ in a liquid metallic system. So, the total number of conduction electrons in this system 
\newpage
\pagestyle{fancy}
\fancyhead{}
\fancyhead[RO]{G M BHUIYAN} 
\fancyhead[LO]{\thepage}
$\hspace{-6mm}$is $N Z$.
The Hamiltonian of the sample may be written as
\begin{eqnarray}
H&=& H_{e} + H_{ee} + H_{ei} + H_{i} + H_{ii} \nonumber \\
 &=& \sum_{i=1}^{NZ} \frac{p_{i}^{2}}{2 M} + \frac{e^{2}}{2}\,\sum_{i\ne j}^{NZ} \frac{1}{|\vec{R}_{i}-\vec{R}_{j}|}+\sum_{i,l} v(|\vec{R}_{i} - \vec{r}_{l})\nonumber \\ 
& & +\sum_{l=1}^{N} \frac{P_{l}^{2}}{2 m} + \frac{1}{2}\,\sum_{l\ne l^{\prime}}^{N} w(|\vec{r}_{l} - \vec{r}_{l^{\prime}}|)
\end{eqnarray}
where $H_{e}$, $H_{ee}$, $H_{ei}$, $H_{i}$, and $H_{ii}$ denote contributions from kinetic energy of electrons, electron-electron interactions, electron-ion interactions,
kinetic energy of ions and ion-ion interactions, respectively. In equation (11) $\{R_{i}\}$ and $\{r_{l}\}$ are electronic and ionic coordination; $\{p_{i}\}$ and $\{P_{l}\}$
electronic and ionic momenta, and, $M$ and $m$ are corresponding masses. $v$ and $w$ denote electron-ion and ion-ion potential energies, respectively.
 
In the canonical ensemble theory the normalized equilibrium probability density $f_{0}^{(N)}$ for a system of homonulear atoms is given by
\begin{eqnarray}
f_{0}^{(N)} (\vec{r}_{1},\cdots,\vec{r}_{N}, \vec{p}_{1},\cdots,\vec{p}_{N})= \frac{\exp[- \beta H (\vec{r}_{1},\cdots,\vec{r}_{N}, \vec{p}_{1},\cdots,\vec{p}_{N})]}
{N! h^{3N} Q_{N}(V,T)}
\end{eqnarray}
where $h$ denotes Planck's constant, and $Q_{N}(V,T)$ the total partition function,
\begin{eqnarray}
Q_{N}(V,T) &=& \text{Tr}e^{-\beta H} \nonumber \\
&=& \frac{1}{N! h^{3N}} \int d\vec{r}_{1} \cdots d\vec{r}_{N} \int d\vec{p}_{1} \cdots d\vec{p}_{N} \,\text{Tr}_{e} e^{-\beta H}
\end{eqnarray}
where Tr$_{e}$ refers complete set of electronic states corresponding to a particular ionic configuration. The motion of ions is very slow relative to the
conduction electrons, so, ions can be treated classically unlike electrons that must be handled quantum mechanically. As classical particle do not obey
uncertainty principle one can integrate over position and momentum independently. The result thus obtained is
\begin{eqnarray}
Q_{N}(V,T) = \frac{1}{N!} \left[\frac{2 \pi m}{\hbar^{2} \beta} \right]^{\frac{3N}{2}}\, Z_{N}(V,T)
\end{eqnarray}
where the configurational partition function
\begin{eqnarray}
Z_{N}(V,T) = \int \cdots \int  d\vec{r}_{1} \cdots d\vec{r}_{N} \exp(-\beta H_{ii}) \left\{ \int \cdots \int  d\vec{R}_{1} \cdots d\vec{R}_{N}\right. \nonumber \\
\left. d\vec{P}_{1} \cdots d\vec{P}_{N}\, e^{-\beta (H_{e} + H_{ee} +H_{ei})} \right\}
\end{eqnarray}
\newpage
\pagestyle{fancy}
\fancyhead{}
\fancyhead[LO]{MICROSCOPIC ORIGIN OF IMMISCIBILITY $\cdots$} 
\fancyhead[RO]{\thepage}
$\hspace{-6mm}$Ions move in the following effective pair potential
\begin{eqnarray}
U_{N} = H_{ii} + F^{\prime}
\end{eqnarray}
where $F^{\prime}$ is the Helmholtz free energy of the conduction electrons in the external potential $H_{ie}$. $F^{\prime}$ can be calculated
by some approximation schemes. Therefore
\begin{eqnarray}
Z_{N} = \int  d\vec{r}_{1} d\vec{r}_{2}\cdots d\vec{r}_{N}\, e^{-\beta U_{N}}\,,
\end{eqnarray}
and the L-body probability density
\begin{equation}
n_{N}^{(L)} = \frac{\int \cdots \int d\vec{r}_{L+1} d\vec{r}_{L+2}\cdots d\vec{r}_{N}\, e^{-\beta U_{N}}}{Z_{N}}
\end{equation}
This is related to the $L$-particle distribution function defined as
\begin{eqnarray}
g^{(L)}(\vec{r}_{1} \vec{r}_{2}\cdots \vec{r}_{N}) &\equiv& \frac{n_{N}^{(L)}}{n^{L}} \nonumber \\
&=&\frac{N!}{n^{L} (n-L)!}\,\frac{\int \cdots \int d\vec{r}_{L+1} \cdots d\vec{r}_{N}\, e^{-\beta U_{N}}}{Z_{N}}
\end{eqnarray}
Now, the two body reduced distribution function stands as
\begin{eqnarray}
g^{(2)}(\vec{r}_{1}, \vec{r}_{2}) &=&\frac{N (N-1)}{n^{2}}\,\frac{\int \cdots \int d\vec{r}_{3} \cdots d\vec{r}_{N}\, e^{-\beta U_{N}}}{Z_{N}}
\end{eqnarray}

For an isotropic liquid
\begin{eqnarray}
 g^{(2)}(\vec{r}_{1}, \vec{r}_{2})=g(|\vec{r}_{2}-\vec{r}_{2}|) = g(r)\,, \nonumber
\end{eqnarray}
which is also known as pair correlation function and is the central idea in most liquid state theories.

Now if it is assumed that the effective interionic potential is pairwise additive in the following way
\begin{equation}
U_{N} = N E(V) + \frac{1}{2}\sum_{i,j} v(r_{ij})\,,
\end{equation}
where $E(V)$ is the volume dependent (but structure independent) part of energy that includes the free energy of electrons, then all
thermodynamic
\newpage
\pagestyle{fancy}
\fancyhead{}
\fancyhead[RO]{G M BHUIYAN} 
\fancyhead[LO]{\thepage}
$\hspace{-6mm}$functions can be expressed in terms of $g(r)$ and the pairpotential of interaction. We note here that equation (20) cannot be solved
analytically even if Eqn. (21) is used. For this mathematical limitation different approximation methods and computer simulation methods
are devised to solve for $g(r)$. But for the hard sphere (HS) potential
\begin{eqnarray}
v_{hs}(r) = 
\begin{cases}
\infty & \text{for}\,\,\,\,\,\,  r< \sigma \\
0 & \text{for} \,\,\,\,\,\,\, r > \sigma
\end{cases}
\end{eqnarray}
$g(r)$ can be evaluated analytically [33,40], here $\sigma$ denotes the hard sphere diameter (HSD). The pair correlation function for a liquid
binary alloys [50] may be expressed as
\begin{eqnarray}
g_{ij}(r) = 1 + \frac{1}{(2 \pi)^{3} \rho \sqrt{C_{i}C_{j}}}\,\int(S_{ij}(q) - \delta_{ij})\,e^{i \vec{q} \cdot \vec{r}}\,d^{3}r
\end{eqnarray}
where $S_{ij}(q)$ is the static structure factors and $q$ the momentum transfer.
\subsection{Thermodynamic perturbation theory}
The thermodynamic perturbation theory proposed by Weeks-Chandler-Andersen (WCA)[51] splits the interionic potential as core and tails
terms
\begin{eqnarray}
 v(r) = v_{core}(r) + v_{tail}(r)
\end{eqnarray}
\begin{figure}[!h]                                                                    
\begin{center}                                                                        
\includegraphics[width=8cm,height=6cm]{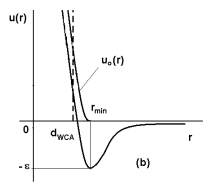}                              
\end{center}                                                                          
\caption{Splitting of the effective pair potential into hard and soft parts.}                                 
\end{figure} 
The core term is related to the HS potential, $v_{hs}$ through the Mayer's cluster expansion in the following way.
\begin{eqnarray}
 f_{\mu}(r) = f_{hs}(r) + \mu \,\Delta f(r) \hspace{1cm} \text{for} \hspace{5mm} 0 \le \mu \le 1 \,,
\end{eqnarray}
where $\mu$ is the coupling parameter, and
\begin{eqnarray}
 \Delta f(r) = f_{core}(r) - f_{hs} = \left[ e^{-\beta v_{core}} - e^{ - \beta v_{hs}} \,\right]
\end{eqnarray}

The Helmholtz free energy can be expanded now as
\begin{eqnarray}
 F_{core}= F_{hs}+ E(V) - \frac{1}{2} k T \rho \, \sigma \,\xi + \mathcal{O}(\xi^{4})
\end{eqnarray}
where 
\begin{eqnarray}
 \xi = \frac{1}{\sigma} \int_{0}^{\infty} B_{hs}(r)\, d\vec{r}
\end{eqnarray}
\newpage
\pagestyle{fancy}
\fancyhead{}
\fancyhead[LO]{MICROSCOPIC ORIGIN OF IMMISCIBILITY $\cdots$} 
\fancyhead[RO]{\thepage}
$\hspace{-6mm}$with the blip function
\begin{eqnarray}
 B(r) = y_{\sigma}(r) \left\{ e^{- \beta v_{core}(r)}-e^{- \beta v_{hs}(r)}\,\right\}\,.
\end{eqnarray}
From equation (27) it is clear that, when $\xi=0$, $F_{core} = F_{hs}+ E(V)$. In the WCA theory hard sphere diameter $\sigma$ is determined following this condition that 
Fourier transform of $B(r)$ that is $B(q)$ vanishes at $r=\sigma$. 
But in WCA theory $r^{2} B(r)$ shows a saw tooth shaped function. If this function is linearzed to have a triangular form one can find an equation [52]
\begin{eqnarray}
\beta v(\sigma) = \ln \left(\frac{- 2 \beta v^{\prime}(\sigma) + X +2}{- \beta v^{\prime}(\sigma) + X +2}\right)\,,
\end{eqnarray}
here, prime indicates the first derivative of the potential energy at $r=\sigma$, and
\begin{eqnarray}
 X= \frac{\sigma/\sigma_{w}}{g_{0}} \left[ \sum_{k=0} \frac{\xi_{k+1} (\eta_{w})}{n!}\,\left(\frac{\sigma}{\sigma_{w}} - 1\right)^{n} - \frac{A \sigma_{w}}{\sigma^{2}}\,(1+ \mu \sigma) \right]
\end{eqnarray}
All symbols are defined in reference [52]. Solution of the transcendental equation (30) yields the effective HSD. The pair correlation function is now
evaluated using this effective HSD.
\newpage
\pagestyle{fancy}
\fancyhead{}
\fancyhead[RO]{G M BHUIYAN} 
\fancyhead[LO]{\thepage}
Andersen et al.[53] proposed a simplest version of the perturbative scheme known as exponential approximation, $$g_{hs}(r)=g_{}\,e^{-v(r)/kT}$$ where
$v(r)$ is the real short range part of the potential; in the present case it is $v_{core}$. We note that this optimized form gives more realistic description
of the pair correlation function.

Now using the perturbation theory one can calculate the free energy of a system per ion as
\begin{eqnarray}
 F= F_{unp}+ 2 \pi \rho \int v_{pert}\, g_{hs} \,d^{3}r 
\end{eqnarray}
where $$F_{unp} = E(V) + F_{hs}= F_{vol} + F_{gas} + F_{hs}$$ and $$v_{pert} = v_{tail}\,.$$
\begin{eqnarray}
 F_{vol}= \frac{1}{32 \pi^{3}} \int_{0}^{\infty} q^{4} \left\{\frac{1}{\epsilon(q)} - 1\right\}\,|v(q)|^{2}\,dq - \frac{Z E_{F}}{3 Y}
\end{eqnarray}
where $Z=C_{1} Z_{1} + C_{2} Z_{2}$; \hspace{0.5cm} $Y = \chi_{elec}/\chi_{F}$, subscripts $elec$ and $F$  denote isothermal compressibilty of the interacting and free electrons,
respectively. The values of $Y$ are obtained from [54].

The electron gas contribution to the free energy per valence in Rydberg unit is
\begin{eqnarray}
 F_{gas}= \frac{2.21}{r_{s}^{2}} - \frac{0.916}{r_{s}} + 0.31\,\ln r_{s} -0.115 
\end{eqnarray}
where 
\begin{eqnarray}
 r_{s}= \left(\frac{3}{4 \pi \rho Z}\right)^{\frac{1}{3}}/a_{0}; \hspace{1cm} \rho= \frac{\rho_{1} \rho_{2}}{C_{1} \rho_{2} + C_{2} \rho_{1}}\,. \nonumber
\end{eqnarray}
\begin{eqnarray}
 F_{hs}&=& \sum_{i} \left[ - \ln \left(\Lambda_{i}^{3} v\right) + \ln C_{i}\right] - \frac{2}{3}\,\left( \frac{5}{3} -y_{1} + y_{2} + y_{3}\right)\nonumber \\
& &+ (3 y_{2} - 2y_{3})/(1-\eta) + \frac{3}{2} \left( 1- y_{1} -y_{2} - \frac{y_{3}}{3}\right)/(1-\eta)^{2}\nonumber \\ & & + (y_{3}-1) \,\ln(1-\eta)
\end{eqnarray}
\newpage
\pagestyle{fancy}
\fancyhead{}
\fancyhead[LO]{MICROSCOPIC ORIGIN OF IMMISCIBILITY $\cdots$} 
\fancyhead[RO]{\thepage}
$\hspace{-6mm}$where $$\Lambda_{i} = \left\{\frac{2 \pi \hbar^{2}}{m_{1}^{C_{1}} m_{2}^{C_{2}} k T}\right\}^{\frac{1}{2}}\,,$$ $$ \eta=\sum_{i} \eta_{i};
 \hspace{1cm} \eta_{i} = \frac{C_{i} \pi \rho_{i} 
\sigma_{ii}^{3}}{6}\,,$$
$$F_{tail}= D \sum_{i,j} C_{i} C_{j} \,M_{ij} \,,$$  $D= 2 \pi \rho\,,$ $$M_{ij}=\int_{\sigma}^{\infty} v_{ij}\,g_{ij}(r)\, r^{2} \,dr\,. $$
Now, the energy of mixing
\begin{eqnarray}
 \Delta F = \Delta F_{vol} + \Delta F_{gas}+\Delta F_{hs}+\Delta F_{tail}
\end{eqnarray}
$\Delta F_{y}$ to be calculated by using equation (8). Now if the experimental densities of the alloy at different concentrations are available, 
and if the difference between calculated density and the experimental
ones exists and significant an excess volume correction to be added with the thermodynamics of mixing as [34]
\begin{eqnarray}
 \Delta F = \Delta F_{vol} + \Delta F_{gas}+\Delta F_{hs}+\Delta F_{tail} +\Delta F_{evc}\,.
\end{eqnarray}
\underline{\large{Enthalpy of alloy} :}\\ \\
Enthalpy of the alloy per ion
\begin{eqnarray}
 H &=& E + p V \nonumber \\
&=& \frac{3}{2} k T + E(V) + \frac{\rho}{2} \sum_{i=1}^{2} \int g_{ij}(r)\,v_{ij}(r) \, d^{3}r + p V
\end{eqnarray}
\underline{\large{Entropy of alloy} :}\\ \\
Within the above perturbation scheme the entropy of alloy (devided by $N k$)
\newpage
\pagestyle{fancy}
\fancyhead{}
\fancyhead[RO]{G M BHUIYAN} 
\fancyhead[LO]{\thepage}
$\hspace{-6mm}$reads [38]
\begin{eqnarray}
 S &=& S_{ref} + S_{tail}\,,\\
S_{ref} &=& S_{id}+ S_{gas} + S_{\eta} + S_{\sigma}\,, \nonumber \\
S_{id} &=& - [ C_{1} \ln C_{1} + C_{2} \ln C_{2}]\,, \nonumber \\
S_{gas} &=& \frac{5}{2} \ln \left[\frac{1}{\rho} \left(\frac{m_{1}^{C_{1}} m_{2}^{C_{2}} k T}{2 \pi \hbar^{2}}\right)^{\frac{3}{2}}\right]\,,\nonumber \\
 S_{\eta} &=& \ln (1-\eta)+\frac{3}{2} [1-(1-\eta)^{-2}]\,,\nonumber\\
S_{\sigma}&=& \left[ \frac{\pi C_{1} C_{2} \rho (\sigma_{11}^{2}-\sigma_{22}^{2}) (1-\eta)^{-2}}{24}\right]\nonumber \\
& & \times \{ 12\, (\sigma_{11}+\sigma{22}) - \pi\, \rho \,[ C_{1} \sigma_{11}^{4} + C_{2} \sigma_{22}^{4}] \}\,, \nonumber
\end{eqnarray}
and
\begin{eqnarray}
 S_{tail}= \frac{1}{k} \left[\left(\frac{\partial F_{tail}}{\partial T}\right)_{V,\rho,\sigma_{ii}} +\sum_{i=1}^{2} \left(\frac{\partial F_{tail}}{\partial \sigma_{ii}}\right)_{V,T}
\left(\frac{\partial \sigma_{ii}}{\partial T}\right)_{V,\rho}\right]\,.
\end{eqnarray}
The temperature dependent HSD as proposed by Protopapas et al. [55] is
\begin{eqnarray}
 \sigma(T) = 1.126\,\sigma_{m} \left\{ 1-0.112 \left(\frac{T}{T_{m}}\right)^{\frac{1}{2}} \right\}.
\end{eqnarray}
 Entropy of mixing therefore stands
\begin{eqnarray}
 \Delta S = \Delta S_{ref} + \Delta S_{tail}
\end{eqnarray}
An alternative way may also be used to evaluate entropy of mixing
\begin{eqnarray}
 \Delta S = \frac{\Delta H - \Delta F}{T}\,.
\end{eqnarray}
\subsection{The Pseudopotential model}
The effective electron-ion interaction between a conduction electron and an ion may be written as  (in atomic unit)[56]
\begin{eqnarray}
 w(r) =\left\{
\begin{array}{lr}
\sum_{m=1}^{2} B_{m}\exp(-r/ma) & \text{if} \,\,\,\, r<R_{c}  \\
 -Z/r & \text{if} \,\,\,\, r > R_{c}\,,
\end{array} \right.
\end{eqnarray}
\newpage
\pagestyle{fancy}
\fancyhead{}
\fancyhead[LO]{MICROSCOPIC ORIGIN OF IMMISCIBILITY $\cdots$} 
\fancyhead[RO]{\thepage}
$\hspace{-6mm}$where  $Z$, $R_{c}$ and $a$ are the effective $s$-electron occupancy number, core radius and, the softness parameter, respectively. $B_{m}$
is the coefficient of expansion which is independent of $r$ but depends explicitly on parameters $Z$, $R_{c}$ and $a$. The pseudopotential theory leads to an
expression for effective interionic potential of an alloy through the energy band structure [33,36],
\begin{eqnarray}
 v_{ij}(r)= \frac{Z_{i} Z_{j}}{r} \,\left[ 1- \frac{2}{\pi} \int dq \, F_{ij}^{(N)}\,\frac{\sin qr}{q}\right] \,.
\end{eqnarray}
where the wave number characteristics
\begin{eqnarray}
 F_{ij}^{(N)} = \left[ \frac{q^{2}}{8 \pi \rho \sqrt{Z_{i} Z_{j}}}\right]^{2} \,w_{i}(q) w_{j}(q)\,\left[1-\frac{1}{\epsilon(q)}\right]\,[1-G(q)]^{-1}
\end{eqnarray}
%
\subsection{Noticeable beckon and phenomenological theory of segregation}
For a condensed state one of the most basic ingredients from which any microscopic description begins is the subatomic interaction or interionic interaction derived from the former one.
This interaction dictates if the alloy would be an ordered or a segregating type. In ordered alloy, the unlike atoms are preferred as nearest neighbours to like atoms,
whereas in segregating alloys like atoms are preferred as nearest neighbours to unlike atoms. But direct identification of like and unlike atoms in the sample 
is very difficult to achieve experimentally. Indirect ways through some probes assigned with interionic interactions, for example, structural data, thermodynamics of mixing 
(viz. energy of mixing, enthalpy of mixing, entropy of mixing), atomic transport properties (viz. coefficient of shear viscosity, diffusion coefficient)
 and electronic transport properties (viz. resistivity) provide good alternative ways. Some of the microscopic parameters used in identifying segregating alloys are \\
(i) downward concavity or positivity of the free energy of mixing vs concentration profile at any or some concentrations, \\
(ii) upward concavity or negativity of the entalpy of mixing profile at any or some concentrations, \\
(iii) upward concavity or negativity of the entropy of mixing profile at any or some concentrations, \\
(iv) order potential $v_{ord} = v_{ij}(r) - \frac{v_{ii} +v_{jj}}{2} > 0$\, around the nearest neighbour 
\newpage
\pagestyle{fancy}
\fancyhead{}
\fancyhead[RO]{G M BHUIYAN} 
\fancyhead[LO]{\thepage}
$\hspace{-6mm}$distance, \\
(v) a strong bending of the viscosity vs concentration curve near the critical concentration at and below the critical temperature.\\

 In the phenomenological theories or empirical methods [26] there are some other parameters also to identify the segregation of alloys, for example, \\
(i) the Warren-Cowly short range order parameter $\alpha > 0$, \\ 
(ii) in the regular solution theory the exchange energy $w > 0$ \,, \\
(iii) the concentration-concentration structure factors in the long wavelength  limit $S_{cc}(0)$ diverges near the critical temperature and, the sharp 
increase happens around the critical concentration.

Some alloys such as $\text{Li-Na, Al-Bi, Al-Sn, Fe-Cu, Cu-Co, Al-Pb, Bi-Zn, Cd-Ga}$, $\text{Ga-Pb, Ga-Hg, Pb-Zn, Pb-Si, and Cu-Pb}$ are well known systems for which 
some segregating properties are measured. So, it is worth pursuing to understand the microscopic origin of segregation from the theoretical point of view
and compare them with the experimental ones.
\subsection{Phenomenological theories in the study of segregating properties}
The Gibbs free energy of mixing ( of a sample of $N$ moles) for binary alloys is
\begin{eqnarray}
 \Delta G = G_{alloy} - \sum_{i=1}^{2} C_{i} G_{i} \,.
\end{eqnarray}
In terms of the partial Gibbs energies $\Delta G_{i}$, one can write
\begin{eqnarray}
 \Delta G = C_{i} \Delta G_{i} + C_{j} \Delta G_{j}\,,
\end{eqnarray}
with $$\Delta G_{i} = R\, T \ln a_{i}\hspace{1cm} (i=1,2)$$ where $a_{i}$ denotes the thermodynamic activity of the $i$-th component.

The stability of a binary mixture is mostly determined by $\Delta G$. Figure 2($i$) shows a schematic diagram for $\Delta G$ denoted by $G_{M}$ as a function of concentration
$C$. Here curve $a$ describes a miscible stable state whereas curve b describes an immiscible unstable state in the concentration range $\Delta C$.
The points P and Q in Figure 2($i$) give compositions of two segregated 
\newpage
\pagestyle{fancy}
\fancyhead{}
\fancyhead[LO]{MICROSCOPIC ORIGIN OF IMMISCIBILITY $\cdots$} 
\fancyhead[RO]{\thepage}
$\hspace{-6mm}$phases. At points $P$ and $Q$ the partial Gibbs energies of the components are
equal, $$\Delta G_{i}(C_{1}) = \Delta G_{i}(C_{2}) \hspace{1cm} (i= A, B). $$ The point of inflexion in the curve b for $T_{2} < T_{c}$ represents the spinodal line.
The critical concentration and critical temperature follow from the following conditions
\begin{eqnarray}
 \left(\frac{\partial^{2} \Delta G}{\partial C^{2}}\right)_{C=x_{c}}=0\,\,\, ; \hspace{1cm} \left(\frac{\partial^{3} \Delta G}{\partial C^{3}}\right)_{C=x_{c}}=0 \nonumber
\end{eqnarray}
at $T=T_{c}$.
\begin{figure}[!h]                                                                    
($i$)\includegraphics[width=6cm,height=6cm]{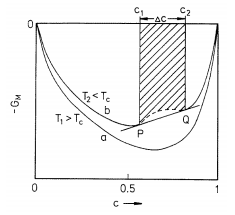}                                                                                                     
($ii$)\includegraphics[width=6cm,height=5.5cm]{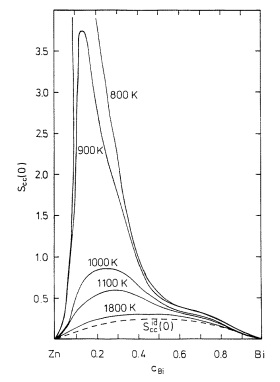}                              
\caption{($i$) A schematic diagram of Gibbs free energy of mixing as a function of concentration and ($ii$) $S_{CC}(0)$ for different temperatures (after Singh and Sommer [26]).}                                 
\end{figure} 
Following the Bhatia-Thronton structure factors [57], which is well known for the concentration-concentration fluctuation in the long wave length limit, one can show
\begin{eqnarray}
S_{CC} (0) = R T \left(\frac{\partial^{2} G_{m}}{\partial C^{2}}\right)^{-1}_{T,p}  \,.\nonumber
\end{eqnarray}

As $$C\longrightarrow x_{c}, \hspace{0.5cm}\text{and}\hspace{0.5cm} T \longrightarrow T_{C},\hspace{0.5cm}S_{CC} (0) \longrightarrow \infty \,.$$ 
This property of $S_{CC}(0)\longrightarrow \infty$ signals the phase separation in a binary mixture. Figure 2($ii$) shows this behaviour. 
Other empirical models used 
\newpage
\pagestyle{fancy}
\fancyhead{}
\fancyhead[RO]{G M BHUIYAN} 
\fancyhead[LO]{\thepage}
$\hspace{-6mm}$in the study of demixing of alloys are quasi-lattice theory [7,26] and the self association model [58]. Using the quasi lattice theory [7,26] it is possible 
to derive the configurational energy and partition function of the alloy. This knowledge later yields an expression for the Gibbs free energy of mixing and thermodynamic activity  
in terms of a free parameter known as interchange energy. The critical properties of segregation can then be obtained from the so called stability conditions
\begin{eqnarray}
 \frac{\partial \ln a_{i}}{\partial C_{i}}=0\,\,\, ; \hspace{1cm} \frac{\partial^{2} \ln a_{i}}{\partial C_{i}^{2}}=0 \,,\nonumber
\end{eqnarray}
for different clusters suggested by the self association model.

\begin{table}
\caption{Critical concentration and critical for demixing liquid alloys.}
\vspace{0.5cm}
\begin{center}
\begin{tabular}{lcccl} 
\hline 
Systems &  &   &  &   \\
A$_{m}$-B$_{n}$ & m & n & x$_{c,A}$ & $\frac{w}{k T_{c}}$ \\
\hline\\
A-B & 1 & 1 & 0.5 & 2.0\\
A$_{2}$-B$_{2}$ & 2 & 2 & 0.5 & 1.0 \\
A$_{4}$-B$_{4}$ & 4 & 4 & 0.5 & 0.5 \\
A-B$_{2}$ & 1 & 2 & 0.74 & 1.457 \\
A$_{2}$-B$_{4}$ & 2 & 4 & 0.74 &  0.728 \\
A-B$_{3}$ & 1 & 3 & 0.84 & 1.244 \\
A$_{2}$-B$_{8}$ & 2 & 8 & 0.89 & 0.562 \\ \hline\\
\end{tabular}
\end{center}
\end{table}
Although this empirical theory presents a good prescription to study the critical properties of segregating alloys, its reliability in predicting
critical properties of real binary alloys is yet to be seen.
\newpage
\pagestyle{fancy}
\fancyhead{}
\fancyhead[LO]{MICROSCOPIC ORIGIN OF IMMISCIBILITY $\cdots$} 
\fancyhead[RO]{\thepage}
%
\section{Results from the microscopic approach}
\subsection{Partial pair potentials and corresponding pair correlation functions}
For any microscopic description of a condensed matter the most fundamental ingredient necessary is the knowledge of interionic potential.
Direct derivation and application of the $N$-body potentials to the study of the physical properties of condensed matter is a too much difficult job to handle theoretically.
In order to avoid this difficult situation one goes for the effective pairpotentials. The term effective indicates that, these potentials take into account the many body effects 
in an average way following indirect routes.
\begin{figure}[!h]                                                                    
\begin{minipage}{4cm} 
\centering                                                                       
\includegraphics[width=4cm,height=3cm]{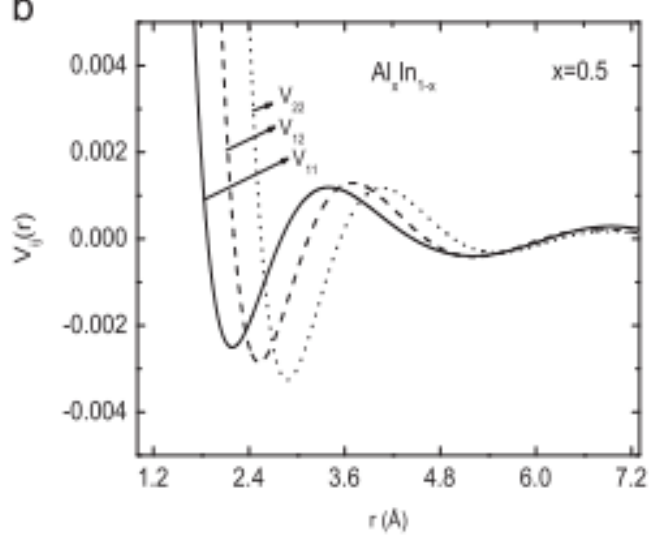} 
\end{minipage}
\begin{minipage}{4cm} 
\centering                             
\includegraphics[width=3cm,height=4cm,angle=270]{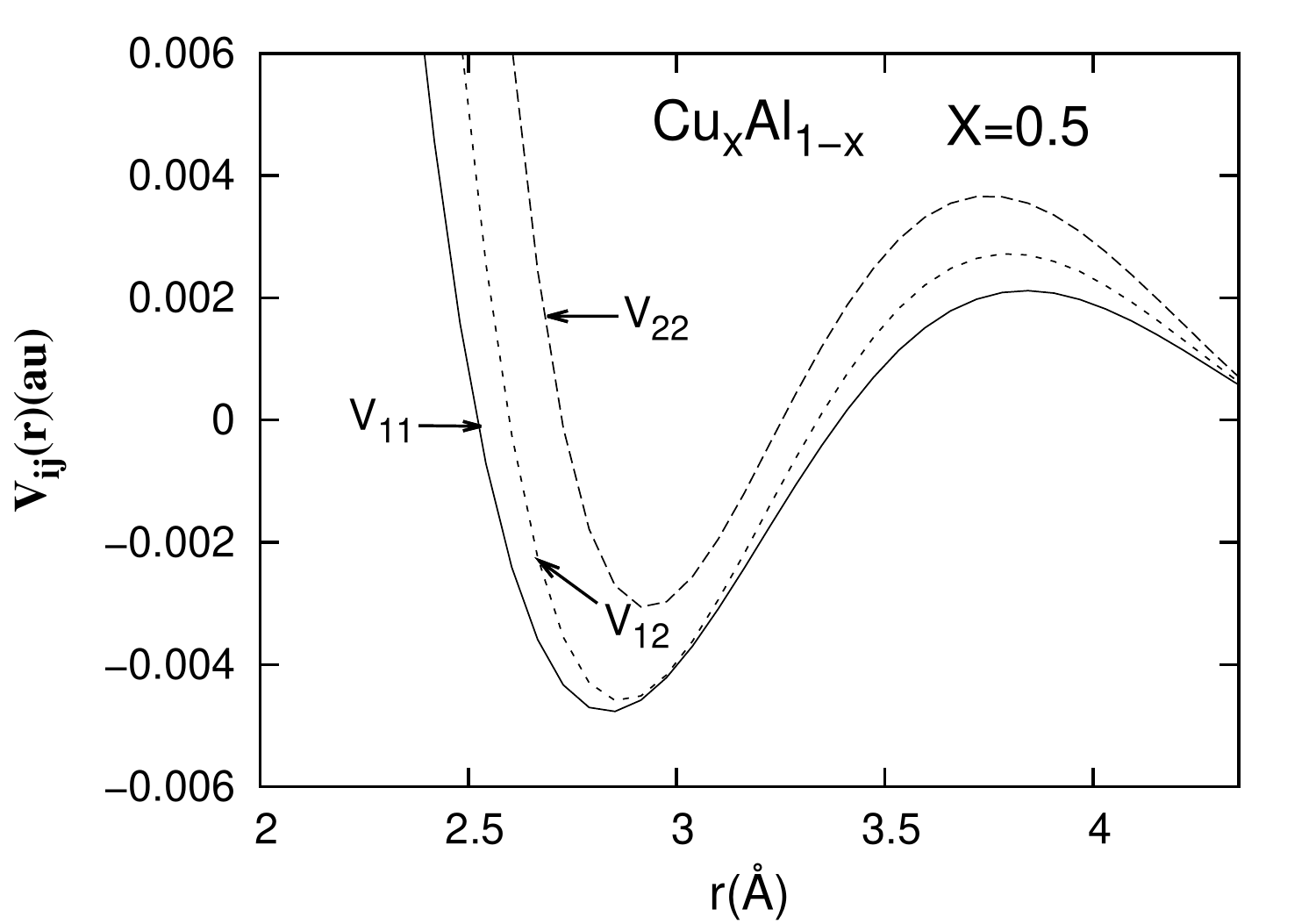} 
\end{minipage} 
\begin{minipage}{4.5cm} 
\centering                            
\includegraphics[width=3.2cm,height=4.5cm,angle=270]{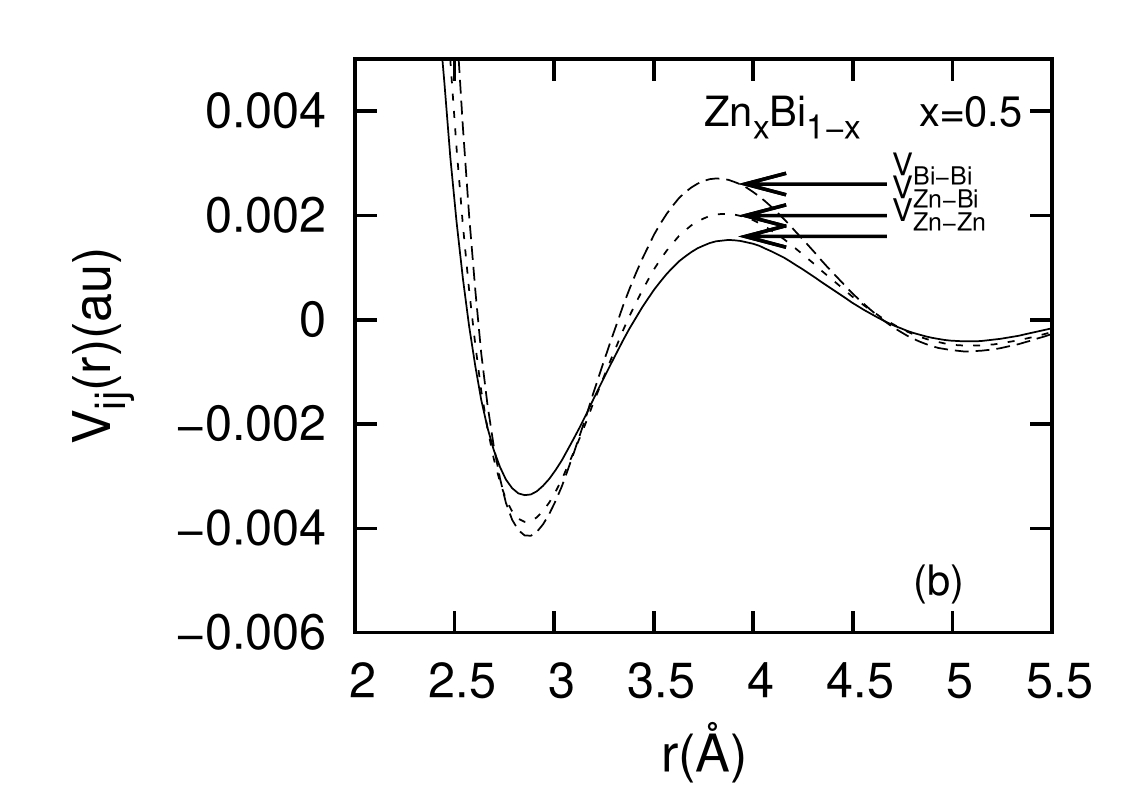}                              
\end{minipage}                                                                          
\caption{ Partial pair potentials for Al$_{x}$In$_{1-x}$, Cu$_{x}$Al$_{1-x}$ and Bi$_{x}$Zn$_{1-x}$ liquid binary alloys For $x=0.5$ (after (from the left) Bhuiyan and coworkers [21, 22, 24] ).}                                 
\end{figure}         
Figure 3 shows the profile of the effective partial pair potentials for an Al-based alloy namely Al$_{x}$In$_{1-x}$. It is seen that partial pair potential $v_{AlAl}(r)$ 
has the sallowest potential well and $v_{InIn}$ the deepest well. That of $v_{AlIn}$ lies in between. It is also seen that the position of the well minima for $v_{AlIn}$ and $v_{InIn}$
shift to large $r$ relative to $v_{AlAl}$. Similar feature is also observed for transition metal segregation alloys ( for example Fe$_{x}$ Cu$_{1-x}$, Cu$_{x}$ Co$_{1-x}$). In case of 
Zn$_{x}$Bi$_{1-x}$ the amount of shift among different partial pair potentials is significantly small. This shifting is largely associated with the difference in the values of
the core radii between individual components of the alloy. We note here that in random alloys $v_{12}$ generally lies between $v_{11}$ and $v_{22}$. But in the case
of compound forming alloys $v_{12}$ goes down the well of the $v_{11}$ or $v_{22}$ whichever has lower value. We note here that in the effective pair potential
calculations Ichimaru-Utshumi dielectric function [59] has been used by Bhuiyan and his group because this function satisfies both compressibility 
\newpage
\pagestyle{fancy}
\fancyhead{}
\fancyhead[RO]{G M BHUIYAN} 
\fancyhead[LO]{\thepage}
$\hspace{-6mm}$sum rule and the short
range correlation function. The BS pseudopotential model has proven to be successful in the studies of liquid structure [60-63], thermodynamic properties [34,35,64,65], 
atomic transport [66-70] and electronic transport properties [71,72] of liquid metals and there alloys.
\begin{figure}[!h]                                                                    
\begin{center}                                                                        
\includegraphics[width=4cm,height=4cm,angle=270]{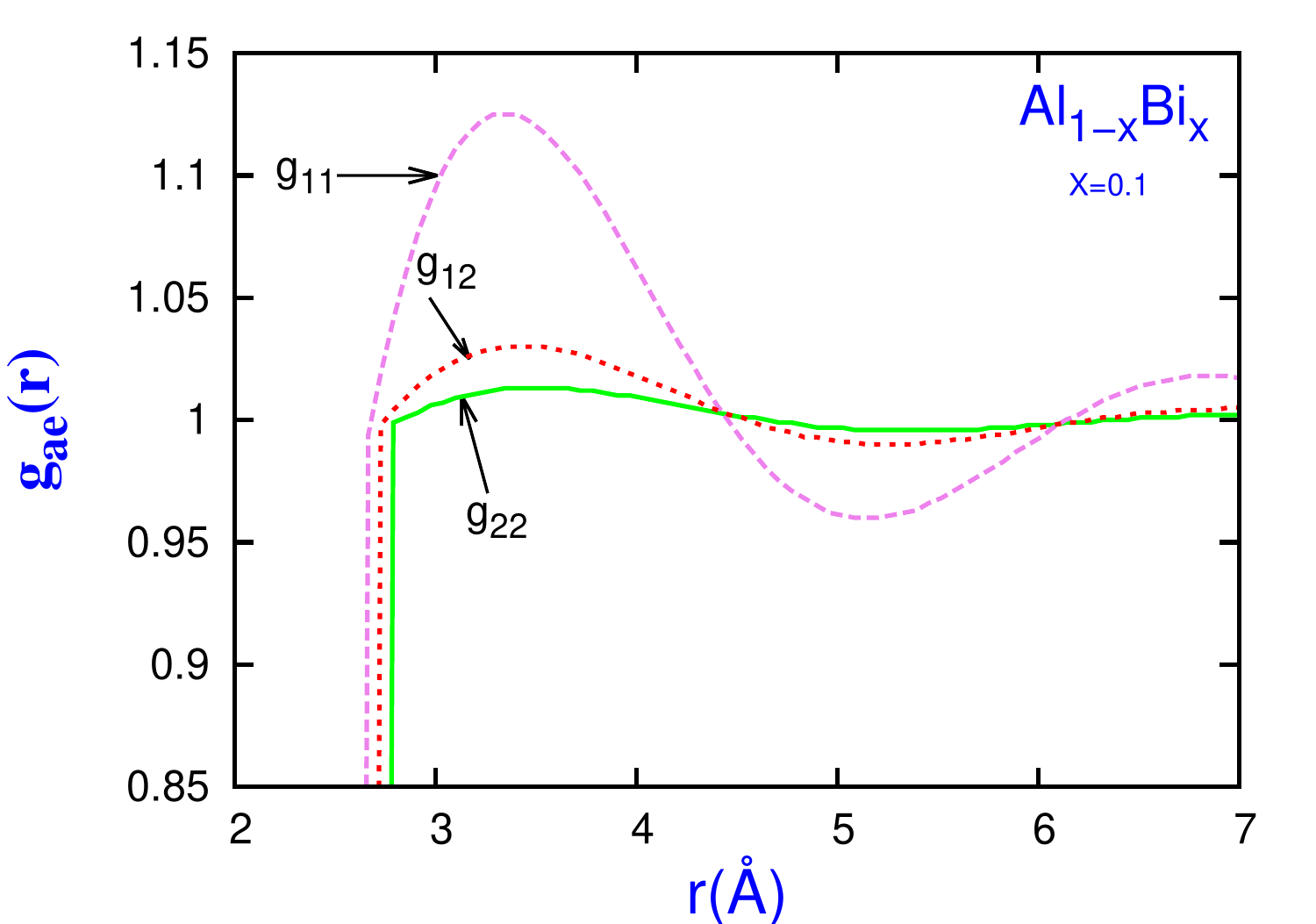}                              
\includegraphics[width=4cm,height=4cm,angle=270]{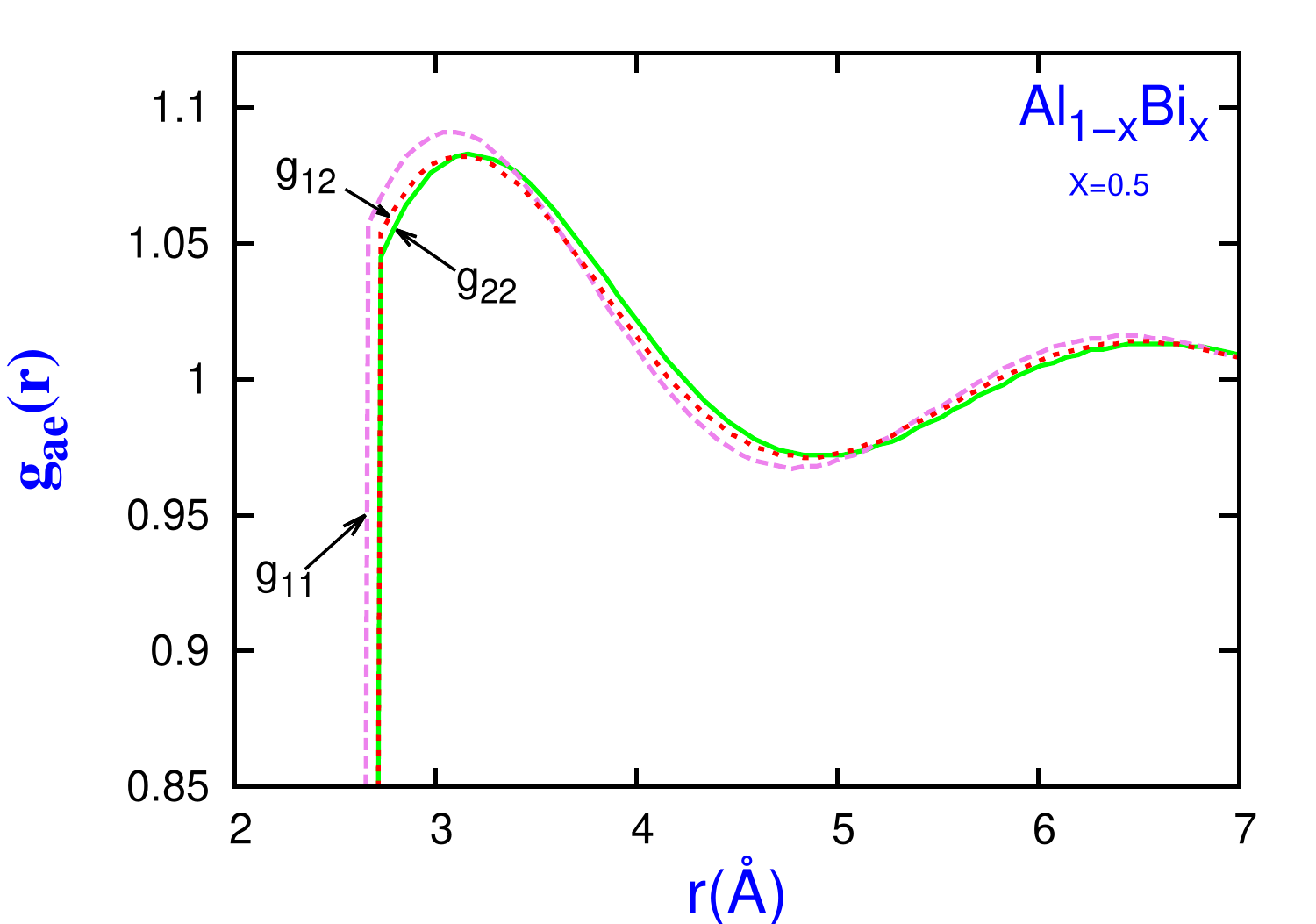}                              
\includegraphics[width=4cm,height=4cm,angle=270]{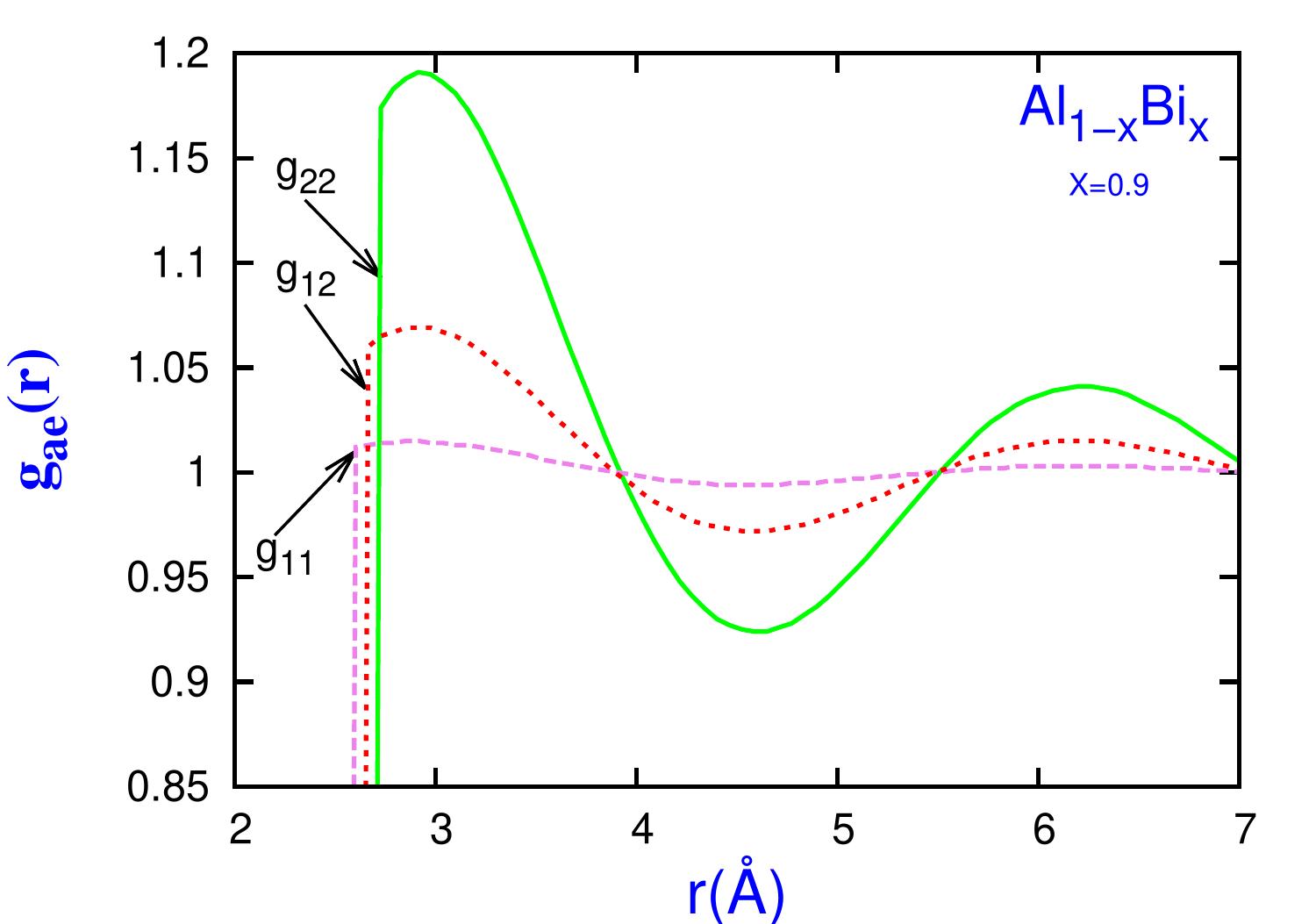}                              
\end{center}                                                                          
\caption{ Partial pair potentials for Bi$_{x}$Zn$_{1-x}$ liquid binary alloys          
         for $x=$0.1, $x=$0.5, $x=$0.9 respectively ( after Kasem et al. [24]).}                                 
\end{figure}                                                                          

The partial pair correlation function, $g(r)$, is related to the partial interionic pair potential through the statistical mechanics[73] (see equation (7)). 
Partial pair correlation functions for three different concentrations are presented in figure 4. In the alloys, rich in component $1$, $g_{11}$ exhibits the largest peak, while the trends become
opposite in alloys rich in component $2$; that is $g_{22}$ shows the largest main peak. But in both cases peak value of $g_{12}$ remains in the middle of $g_{11}(r)$ and $g_{22}(r)$.
The physical significance of $g(r)$ is that, it gives a measure of the probability of finding the number of nearest neighbours at a distance of the peak from the ion
located at the origin. Thus the area under the principal oscillation provides the coordination number, a characteristic feature of the condensed matter.
Advantage of it is that, $g(r)$ can also be derived from the $X$-ray or neutron diffraction data through the Fourier transformation, and directly from the computer simulation experiment.
In the theoretical study of liquid metals it plays the central role in describing thermodynamic properties.

\subsection{Energy of mixing}
The free energy of mixing and its effects on the critical properties of segregation are described for different alloys below.
\newpage
\pagestyle{fancy}
\fancyhead{}
\fancyhead[LO]{MICROSCOPIC ORIGIN OF IMMISCIBILITY $\cdots$} 
\fancyhead[RO]{\thepage}
{(a) \it Li$_{x}$Na$_{1-x}$ liquid binary alloys:}\\

The first attempt to estimate the energy of mixing theoretically for Li$_{1-x}$Na$_{x}$ liquid binary alloys from a microscopic approach was made by Tamaki [74]. He 
also attempted to relate effective pairpotential between ions with the immiscibility of the segregating alloys (see Figure 5(a)).
He was Stroud who made an attempt systematically for the first time to understand the segregating properties such as critical concentration $x_{c}$ and critical temperature $T_{c}$
of Li$_{1-x}$Na$_{x}$ liquid binary alloys using a microscopic theoretical approach [23]. He employed there
the electronic theory of metal based on the empty core model [75], statistical mechanics and the Gibbs-Bogoliubov variational scheme [76] in order to calculate the free energy of mixing.
\begin{figure}[!h]
\centering
(a)\includegraphics[width=5.5cm,height=5cm]{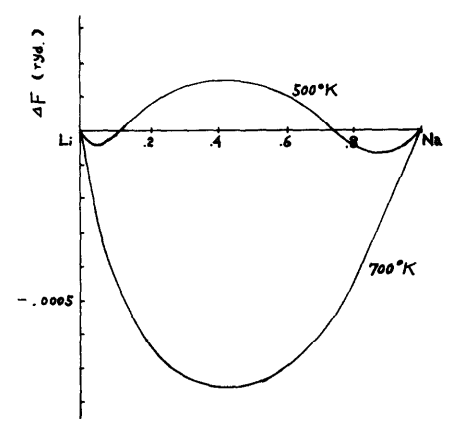}
(b)\includegraphics[width=5.5cm,height=5cm]{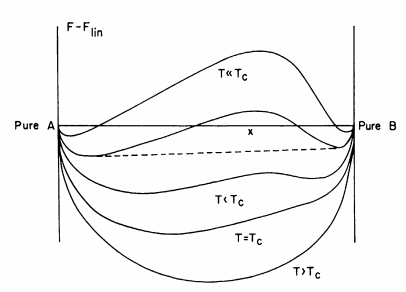}
\caption{ Energy of mixing as a function of $x$ for Li$_{x}$Na$_{1-x}$ liquid binary alloys (after (a) Tamaki [72], (b) Stroud [20])}  
\end{figure}
Figure 5(b) shows a schematic diagram, presented by Stroud in [23], for the $\Delta F$ as a function of concentration for different temperatures. For $T > T_{c}$ the energy of mixing profile 
are concave upward for all concentrations, which manifests complete miscibility (i.e. alloy is stable against segregation) at any concentration. But for $T < T_{c}$ 
the profile becomes concave downward at some concentrations which indicates segregation of the alloy. The temperature at which spinodal points $P$ and $Q$ coincides is
called the critical temperature $T_{c}$, and the concentration at which it happens is called the critical concentration $x_{c}$. Here, in the calculation the Hubbard type
dielectric function [77] is used.
The critical concentration for Li$_{x}$Na$_{1-x}$ segregating alloy was found to be x$_{c}$ =0.7, but the predicted critical temperature was overestimated by one third
\newpage
\pagestyle{fancy}
\fancyhead{}
\fancyhead[RO]{G M BHUIYAN} 
\fancyhead[LO]{\thepage}
$\hspace{-6mm}$[23].\\
\\
{(b) \it Al$_{x}$In$_{1-x}$ liquid binary alloys:}\\

This Al$_{x}$In$_{1-x}$ alloy is formed by the elemental metals Al and In. These elements belong to the less simple polyvalent metals. Al based alloys are known
 to be good candidates for a new advanced anti-friction materials. The input values such as potential parameters $R_{c}$, $a$ and $Z$ along with number density, $\rho$,
 for Al and In, and also for some other elements are shown in Table 2.
\begin{table}
\caption{Potential parameters and densities used for elements that formed different alloys under study are listed.}
\vspace{0.5cm}
\begin{center}
\begin{tabular}{lcccl} 
\hline 
Elements & $\rho\,\,(\AA^{-3})$ & $R_{c}\,(au)$ & $a\,(au)$ & $Z$ \\
\hline\\
Al & 0.0517 & 1.91 & 0.30 & 3.0\\
In & 0.0342& 1.32 & 0.29 & 3.0 \\
Bi & 0.0289 & 1.49 & 0.36 (0.35) & 3 (5) \\
Fe & 0.0756 & 1.425 & 0.33 & 1.5 \\
Co & 0.0787 & 1.325 & 0.27 &  1.5 \\
Cu & 0.0760 & 1.510 & 0.44 & 1.5 \\
 \hline\\
\end{tabular}
\end{center}
\end{table}
\begin{figure}[!h]
\centering
(a)\includegraphics[width=6cm,height=5cm]{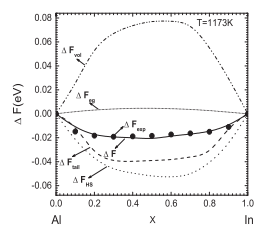}
(b)\includegraphics[width=6cm,height=5cm]{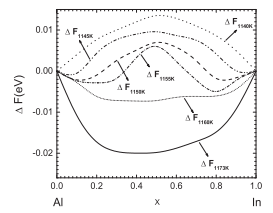}
\caption{ Energy of mixing as a function of $x$ for Al$_{x}$In$_{1-x}$ liquid binary alloys 
        (a) breakdown details at T=1173 K, (b) Temperature dependence.(after Faruk and Bhuiyan [21])} 
\end{figure}

Faruk and Bhuiyan[21] studied the segregating properties of Al$_{x}$In$_{1-x}$ liquid binary alloys by using the electronic theory of metals 
(first principle calculations) along with the statistical mechanics and perturbative approach. Initially, they justified the appropriateness of 
the potential parameters by calculating static structure factors of the elemental liquids at a thermodynamic state at which experimental data are available [78].
Figure 6(a) shows the breakdown details of energy of mixing, $\Delta F$, at $T=$ 1173 K. It is noticed that the HS contribution to the energy of mixing is negative
and values are the lowest among all other contributions across the whole range of concentrations. The tail part contribution is also negative across the concentration range
and values are the second lowest among all others. Contribution of the electron gas, $\Delta F_{eg}$, is positive for the full 
concentration range and values 
\newpage
\pagestyle{fancy}
\fancyhead{}
\fancyhead[LO]{MICROSCOPIC ORIGIN OF IMMISCIBILITY $\cdots$} 
\fancyhead[RO]{\thepage}
$\hspace{-6mm}$are very close to zero. The volume dependent (i.e. structure independent) part of the energy of mixing, $\Delta F_{vol}$, 
due to electron ion interaction is positive and large across the full range of concentration. The combined effect of all contributions, $\Delta F$, 
agree well with the corresponding experimental data [79]. This signifies the accuracy of the approach for the 
study of energy of mixing at different temperatures.

The temperature dependent energy of mixing for Al$_{x}$In$_{1-x}$ liquid binary alloys for different concentrations are illustrated in figure 6(b). As
temperature is decreased from 1173 K, $\Delta F$ increases gradually and at 1155 K becomes partially positive and partially negative.
Further lowering of temperature increase the miscibility gap and at 1140 K the concentration gap span the whole range of concentration. A careful observation
finds the first downward concavity or positivity of $\Delta F$ at 1160 K, and the concentration at which it happens is $x=0.5$. So, the predicted critical temperature and 
concentration for Al$_{x}$In$_{1-x}$ segregating alloys are $T_{c}=1160$ K and $x_{c}=0.5$, respectively. The experimental work by Campbell et al.[80], and Campbell and
Wagemann [9] report a critical temperature of 1220 K. whereas Predel [1] reports 1100 K for $_{x}$In$_{1-x}$ liquid binary alloys. 
Differential thermal analysis by Sommer et al. [81] reports $T_{c}=$ 1112 K.
The average of these scattered experimental data is 1144 K which is close to the theoretical prediction of Faruk and Bhuiyan[21]. 
The experimental critical concentration [10,1] is $x_{c}=$ 0.5 which is exactly the same as that of theoretical prediction 
\newpage
\pagestyle{fancy}
\fancyhead{}
\fancyhead[RO]{G M BHUIYAN} 
\fancyhead[LO]{\thepage}
$\hspace{-6mm}$[21]. But the experimental
data reported in [9] is 0.34 which largely deviates from 0.5. \\ \\
%
\begin{figure}[!h]
\centering
(a)\includegraphics[width=5cm,height=6cm,angle=270]{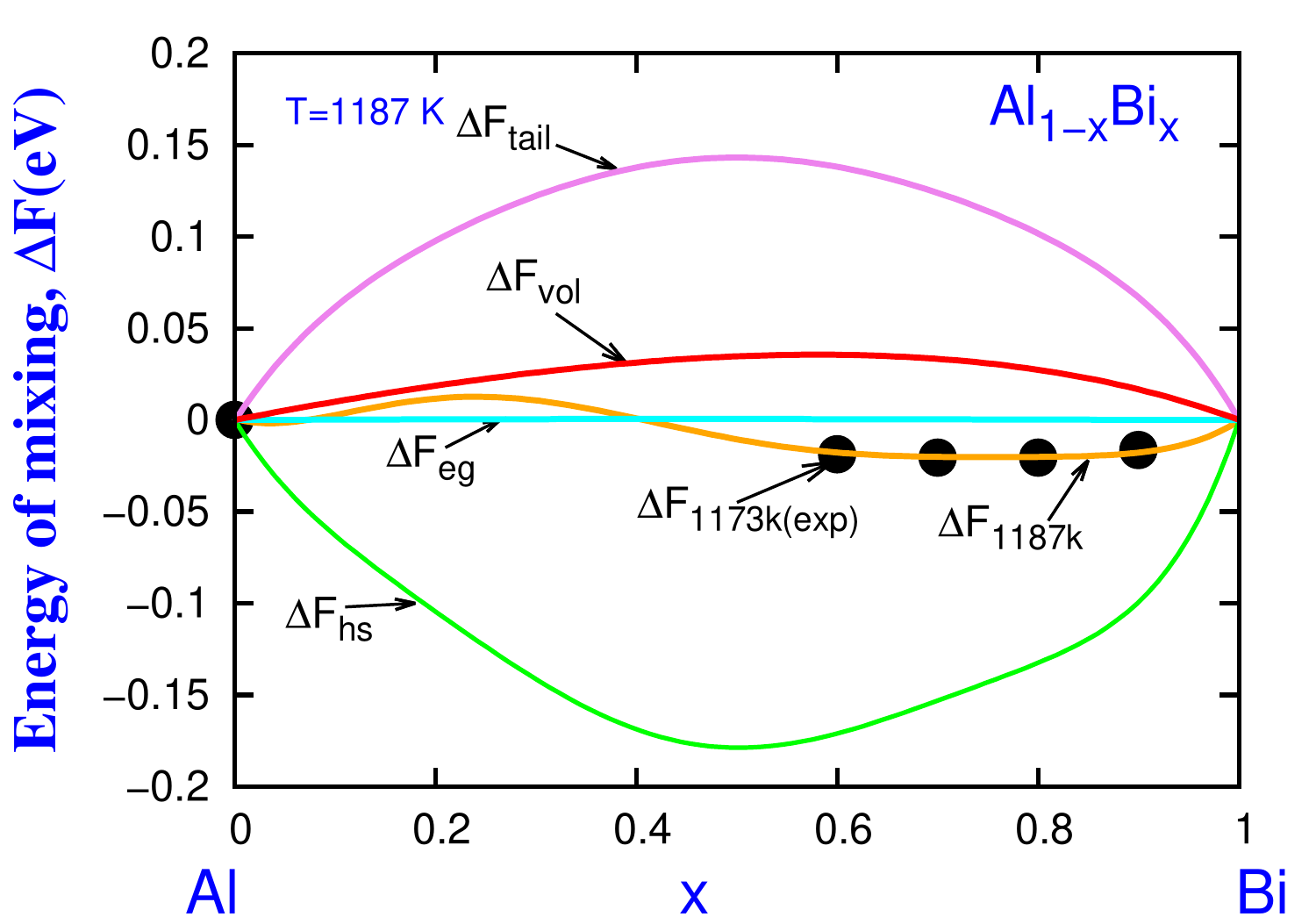}
(b)\includegraphics[width=5cm,height=6cm,angle=270]{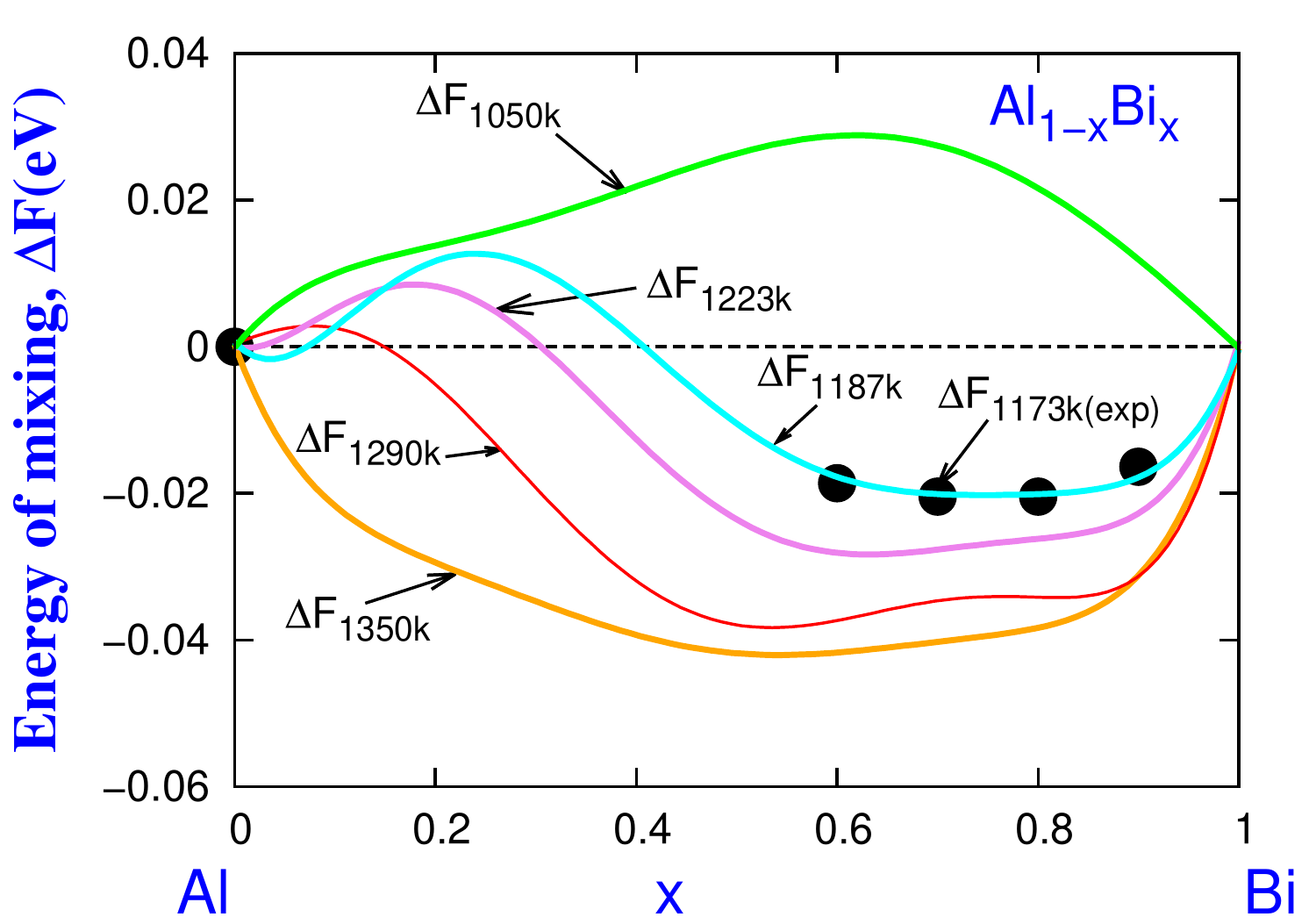}
\caption{ Energy of mixing as a function of $x$ for Bi$_{x}$Al$_{1-x}$ liquid binary alloys 
        (a) breakdown details at T=1187 K, (b) Temperature dependence (after Abbas et al. [25]).} 
\end{figure}
{(c) \it Bi$_{x}$Al$_{1-x}$ liquid binary alloys:}\\

Bi$_{x}$Al$_{1-x}$ liquid binary alloy is formed by two elements Al and Bi which belong to group IIIB and VB in the periodic table, respectively.
The melting points of Al and Bi are 933 and 544 K, respectively; the corresponding densities are 2.375 and 9.78 gm cm$^{-3}$. Al is a trivalent and 
Bi is a pentavalent metal. The atomic radii of Al and Bi are 1.82 and 1.63 \AA, respectively. The large mismatch in their physical properties makes
this alloy interesting to study theoretically.

Figure 7(a) illustrates the breakdown details of the energy of mixing at $T=$1187 K at which some experimental data [79] for $\Delta F$ are available in the literature.
The HS contribution to the energy of mixing is negative for the whole concentration range as is found for Al$_{x}$In$_{1-x}$ liquid binary alloys. But unlike
Al$_{x}$In$_{1-x}$, the tail part contribution, $\Delta F_{tail}$, of Bi$_{x}$Al$_{1-x}$  alloys is positive for all concentrations with a maximum near 
equiatomic concentration. The volume dependent part, $\Delta F_{vol}$, in this case, is positive but the magnitudes are much lower than that of $\Delta F_{tail}$.
The electron gas contribution, $\Delta F_{eg}$, is nearly zero as for Al$_{x}$In$_{1-x}$. The combined effect of all contributions to the free energy, however,
agrees well with the experimental results at $T=$ 
\newpage
\pagestyle{fancy}
\fancyhead{}
\fancyhead[LO]{MICROSCOPIC ORIGIN OF IMMISCIBILITY $\cdots$} 
\fancyhead[RO]{\thepage}
$\hspace{-6mm}$1173 K available in the literature [79].

Figure 7(b) shows the energy of mixing for Bi$_{x}$Al$_{1-x}$  liquid binary alloys for different temperatures. It appears that the alloy exhibits a 
complete miscibility at 1350 K, and immiscibility for all concentrations at 1050 K. But at $T=$ 1290 K, $\Delta F$ shows a partial positivity with concavity downward near $x=$ 0.15.
Further decrease of temperature gradually enhances the concentration gap. As the concavity downward (or positivity) of $\Delta F$ manifests onset of segregation, one can conclude that
the predicted critical concentration is $x_{c}=0.15$, and critical temperature $T_{c}=$ 1290 K, while the corresponding experimental values are $x_{c}=0.19$ [82] and $T_{c}=1310$ [82,83].\\ \\
%
\begin{figure}[!h]
\centering
(a)\includegraphics[width=5cm,height=6cm,angle=270]{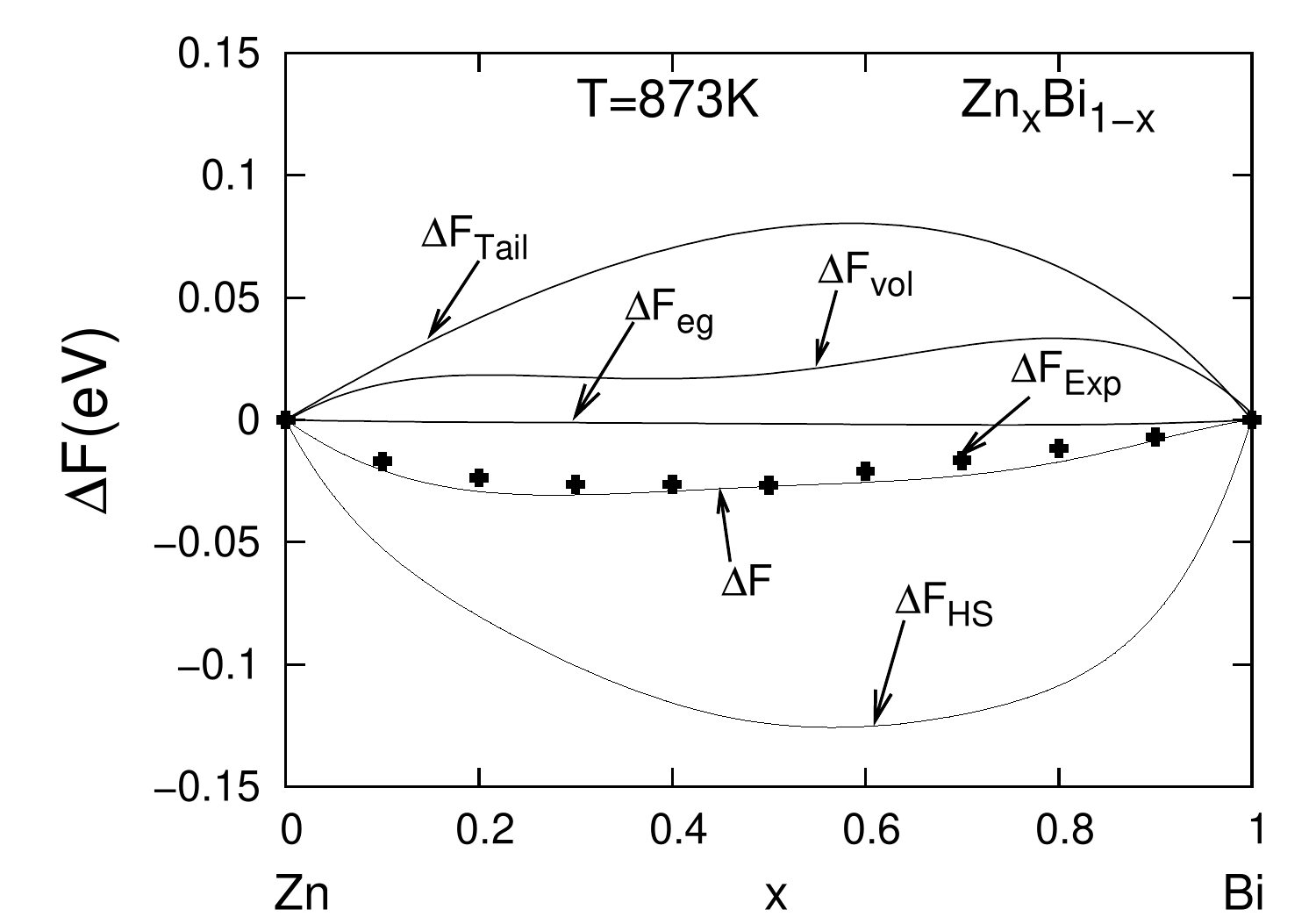}
(b)\includegraphics[width=5cm,height=6cm,angle=270]{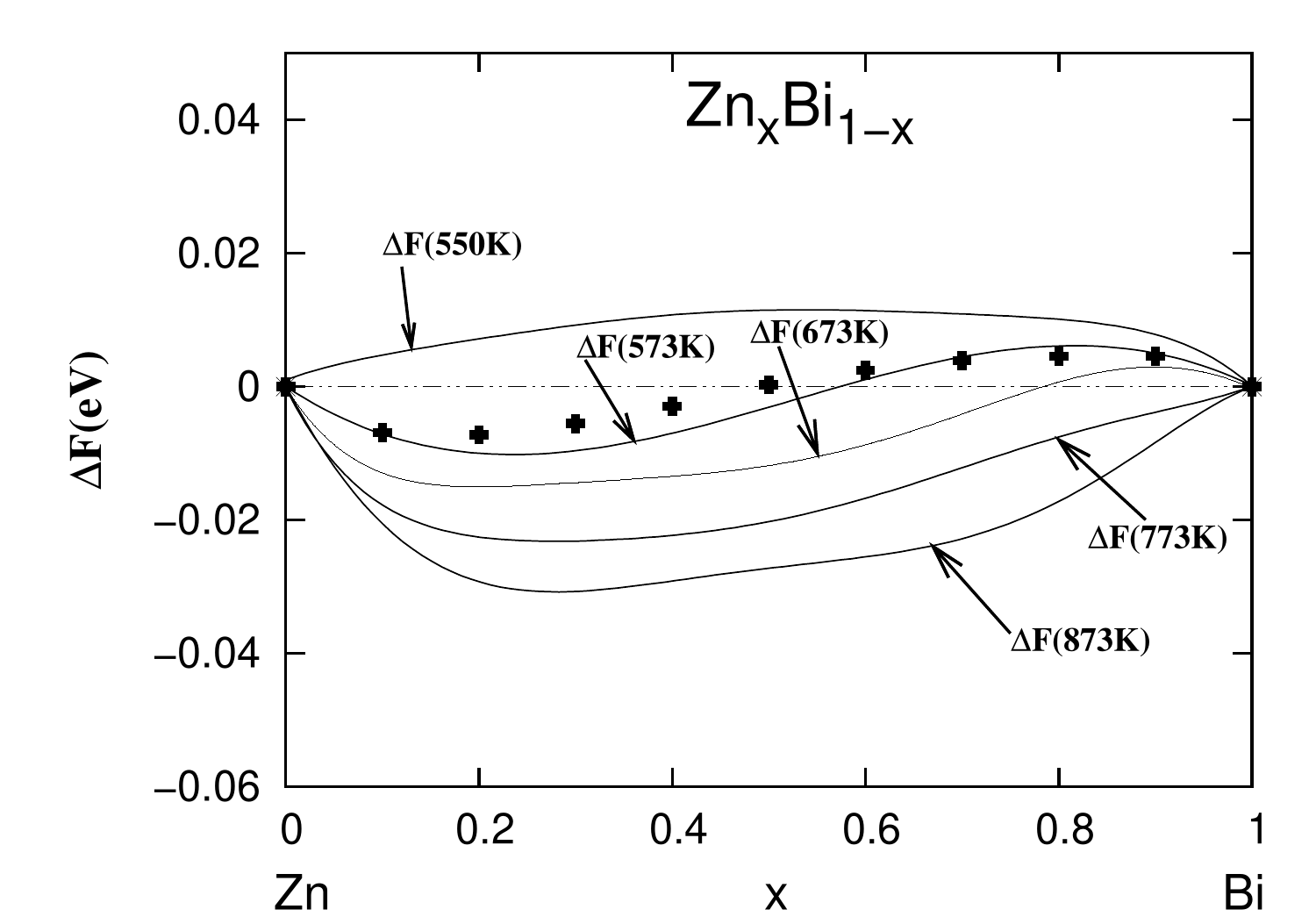}
\caption{ Energy of mixing as a function of $x$ for Bi$_{x}$Zn$_{1-x}$ liquid binary alloys 
        (a) breakdown details at T=873 K, (b) Temperature dependence (after Kasem et al. [24].} 
\end{figure}
{(d) \it Zn$_{x}$Bi$_{1-x}$ liquid binary alloys:} \\

Figure 8(a) shows the free energy of mixing for liquid Zn$_{x}$Bi$_{1-x}$ alloys at 873 K [24]. $\Delta F_{hs}$ is negative for Zn$_{x}$Bi$_{1-x}$ liquid binary alloys 
for the whole range of concentration, this trend is similar to that of the previous alloys, and having the smallest values relative to the other components for each concentration.
In this case $\Delta F$ is asymmetric in nature where the minimum value is found to be around $x=$ 0.6 which is located in the Bi rich alloys. $\Delta F_{tail}$ and $\Delta F_{vol}$
contributions are positive for the full range of concentrations, but $\Delta F_{tail}$ shows the larger values than that of $\Delta F_{vol}$.  $\Delta F_{eg}$ contribution is nearly zero 
as other segregating alloys under consideration of 
\newpage
\pagestyle{fancy}
\fancyhead{}
\fancyhead[RO]{G M BHUIYAN} 
\fancyhead[LO]{\thepage}
$\hspace{-6mm}$[21]this review article. The total energy of mixing, however, matches well with corresponding experimental data [79].

Temperature dependence of $\Delta F$ are illustrated in figure 8(b) It is noticed that at $T=$773 K and higher temperatures $\Delta F$ is negative for all concentrations.
This nature indicates that the alloy is completely miscible in the regime of the above thermodynamic states. But at a lower temperature $T=$ 673 K, $\Delta F$ 
becomes positive i.e. concave downward for some concentrations and negative for others. When temperature is lowered further miscibility gap increases gradually as previous systems 
and cover the whole concentration range at 550 K. From figure it appears that the critical concentration is $x_{c}=0.9$ and the critical temperature $T_{c}=773$ K.
The experimental value for $x_{c}$ is 0.83 [16,17], the critical concentration found theoretically by Stroud [36] and Karlhauber et al [84] (from quasi lattice theory)
was $x_{c}$=0.75 and 0.87, respectively. Experimental critical temperatures are 856 K [16] and 878 K [17], and a theoretical study shows 438 K [36].\\ \\
%
\begin{figure}[!h]
\centering
(a)\includegraphics[width=6cm,height=5cm]{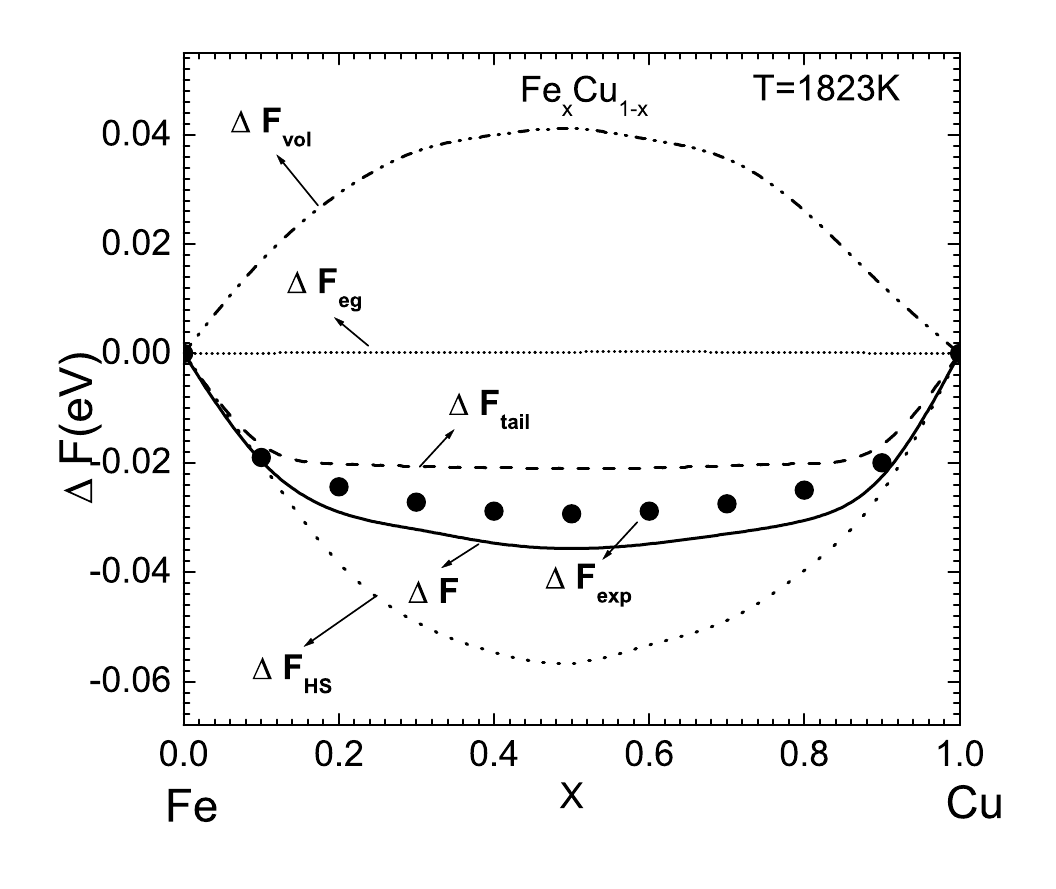}
(b)\includegraphics[width=6cm,height=5cm]{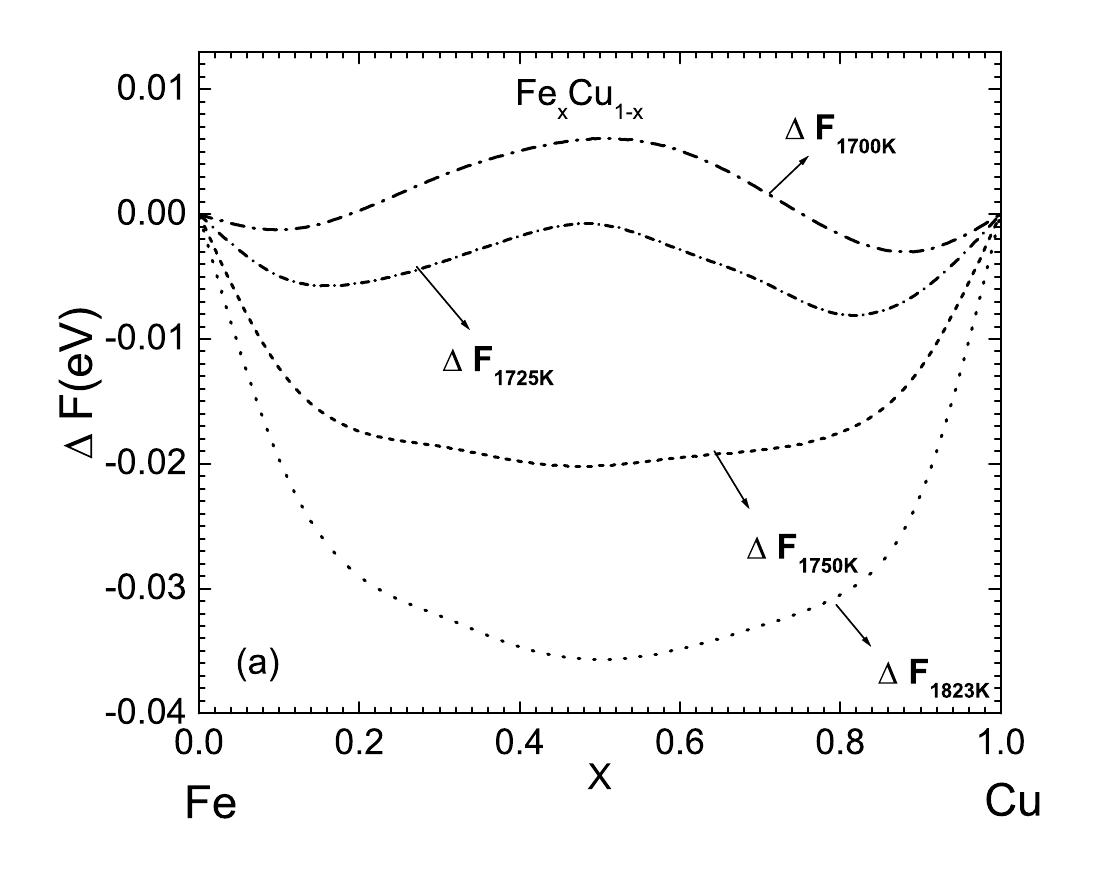}
\caption{ Energy of mixing as a function of $x$ for Fe$_{x}$Cu$_{1-x}$ liquid binary alloys 
        (a) breakdown details at T=1823 K, (b) Temperature dependence (After Faruk et al. [22]).} 
\end{figure}
{(e) \it Fe$_{x}$Cu$_{1-x}$ liquid binary alloys:}\\

Figure 9(a) shows that the HS contribution to the free energy of mixing for Fe$_{x}$Cu$_{1-x}$ is negative for all concentrations at $T=1823$ K [22]. Here the tail part contribution to the 
energy of mixing is negative for all concentrations unlike other segregating alloys. The volume dependent term $\Delta F_{vol}$ is 
\newpage
\pagestyle{fancy}
\fancyhead{}
\fancyhead[LO]{MICROSCOPIC ORIGIN OF IMMISCIBILITY $\cdots$} 
\fancyhead[RO]{\thepage}
$\hspace{-6mm}$positive for the whole 
concentration range, and the electron gas
contribution $\Delta F_{eg}$ is almost zero as for all others discussed above. The total energy of mixing obtained summing all four contributions is negative
for all concentrations and the agreement with available measured data [79] is fairly good. At $T=1823$ K the alloys remain miscible across the full concentration range.
As temperature is lowered to 100 K, $\Delta F$ becomes partially positive around equiatomic concentration where the concavity is downward, and the other part of energy 
of mixing remains negative with upward concavity.
The critical temperature thus found was $T_{c}= 1750 K$ and the critical concentration found was $x_{c}=$ 0.5. The experimental values reported by different authors for
$x_{c}$ are 0.56 [13], 0.538 [14] and 0.538 [15], and the corresponding experimental data for $T_{c}$ are 1696, 1704 K and 1694 K, respectively. \\\\
%
\begin{figure}[!h]
\centering
\includegraphics[width=8cm,height=6cm]{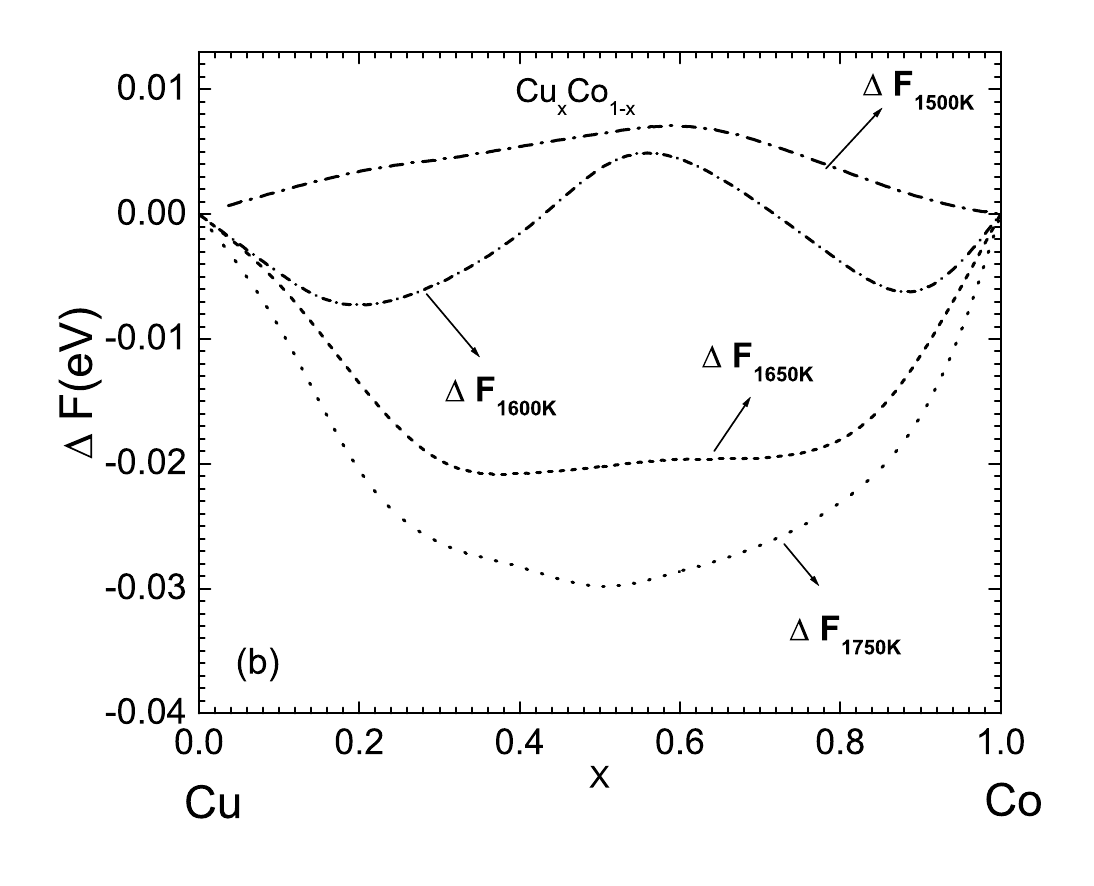}
\caption{ Temperature dependence of energy of mixing as a function of $x$ for Cu$_{x}$Co$_{1-x}$ liquid binary alloys (after Faruk et al. [22]).} 
\end{figure}
{(f) \it Co$_{x}$Cu$_{1-x}$ liquid binary alloys:}\\

In this case the behaviour of various contributions to $\Delta F$ is found to be similar to that of Fe$_{x}$Cu$_{1-x}$ [22]. But, in the immiscible state $\Delta F$ shows (Figure 10)
 an asymmetric feature 
with a value of critical concentration $x_{c}=$ 0.58 and critical temperature $T_{c}=$ 1650 K. The corresponding experimental values are $x_{c}=$ 0.53 and $T_{c}=$ 1547 K [12].

Bhuiyan and coworkers carefully investigated why $\Delta F$ varies with 
\newpage
\pagestyle{fancy}
\fancyhead{}
\fancyhead[RO]{G M BHUIYAN} 
\fancyhead[LO]{\thepage}
$\hspace{-6mm}$temperature. They have found that $\Delta F_{hs}$ and $\Delta F_{tail}$ 
are sensitive to T and are mostly responsible for the variation. While $\Delta F_{vol}$ and $\Delta F_{eg}$ are not sensitive to T at all. The sensitivity arises, in this case,
due to the alteration of $\sigma$ and consequently $g_{hs}(r)$, with the change of $T$.
%
\begin{figure}[!h]
\centering
\includegraphics[width=6cm,height=8cm,angle=270]{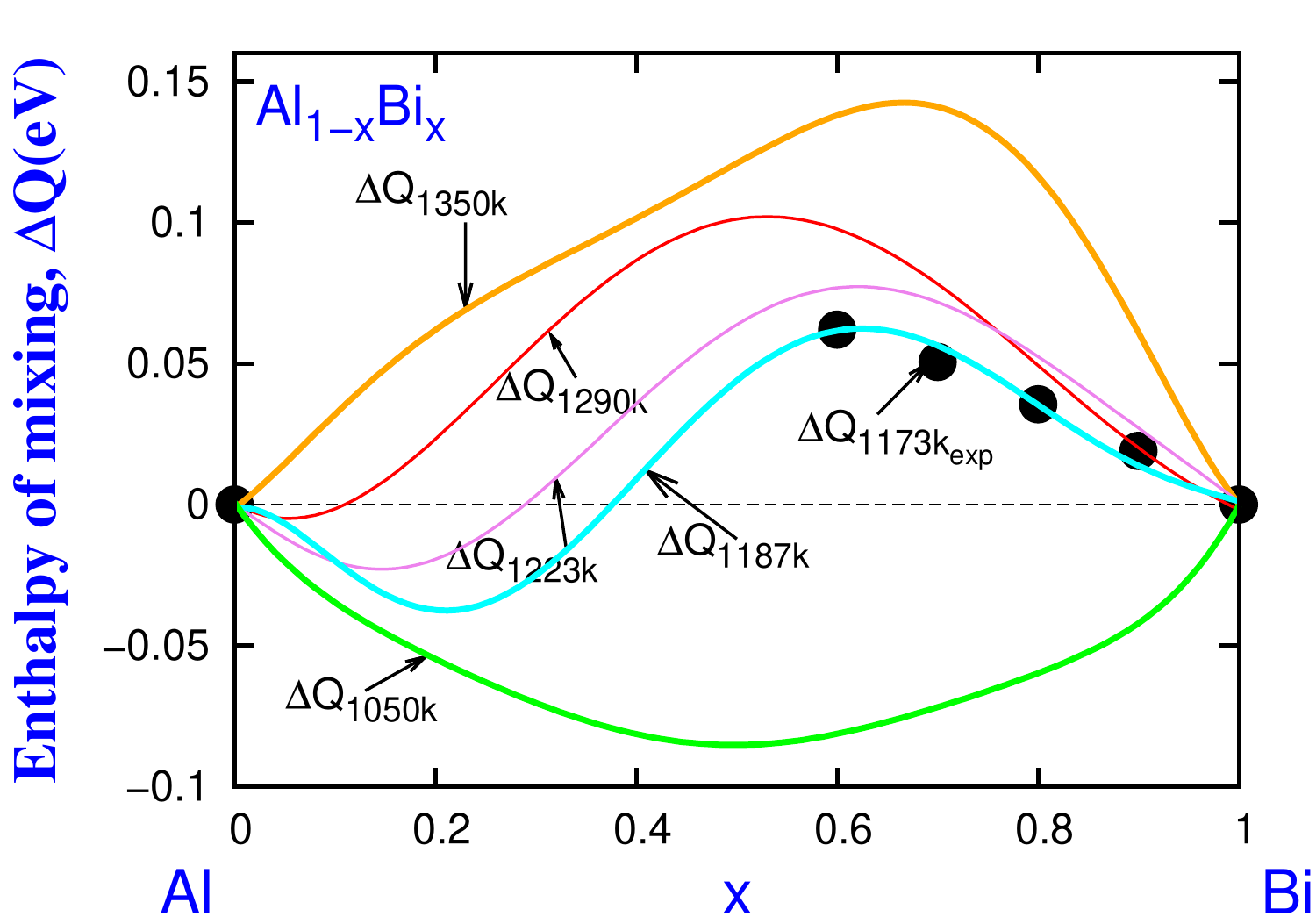}
\caption{ Temperature dependence of enthalpy of mixing as a function of $x$ for Bi$_{x}$Al$_{1-x}$ liquid binary alloys (after Fysol et al. [25]).} 
\end{figure}
\subsection{Enthalpy of mixing}
{(a) \it Bi$_{x}$Al$_{1-x}$ liquid binary alloys:}\\

The enthalpy of mixing, $\Delta H$, are used as a probe to study the critical properties of Bi$_{x}$Al$_{1-x}$ liquid binary alloys. Figure 11 demonstrates that calculated
values of enthalpy of mixing agree in an excellent way with available experimental data for miscible alloys at 1187 K [79]. However, figure also show that the trends of $\Delta H$ as
a function of concentration is just opposite like a mirror reflection to that of free energy of mixing discussed above. That is at 1350 K $\Delta H$ is positive, and 
at 1050 K it is negative for the full concentration range, while at the same thermodynamic states $\Delta F$ shows negative and positive values, respectively. Figure also shows that,
at about 1290 K, $\Delta H$ exhibits negative (i.e. concave upward) at low values of $x$ and positive (i.e. concave downward) for the rest. That means segregation of the alloy begins at 1290 K which is exactly same as found from $\Delta F$ [25].
But in the 
\newpage
\pagestyle{fancy}
\fancyhead{}
\fancyhead[LO]{MICROSCOPIC ORIGIN OF IMMISCIBILITY $\cdots$} 
\fancyhead[RO]{\thepage}
$\hspace{-6mm}$case of enthalpy of mixing the critical concentration is found to be somewhat smaller than predicted by the energy of mixing [25].
\subsection{Enropy of mixing}
It is interesting to see how another static magnitude the entropy of mixing describes the critical properties of segregation for different alloys.\\ \\
\begin{figure}[!h]
\centering
(a)\includegraphics[width=6cm,height=6cm,angle=270]{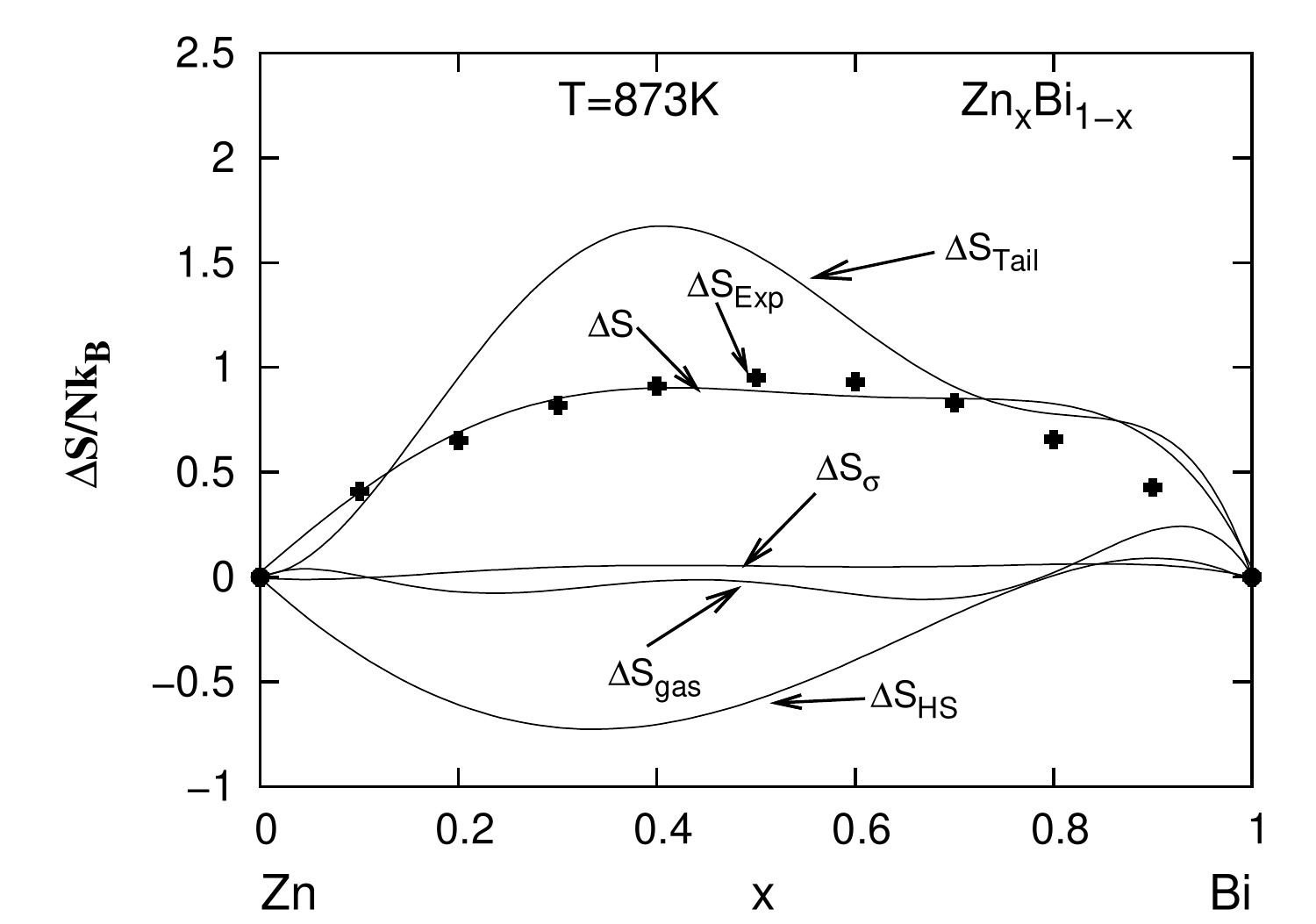}
(b)\includegraphics[width=6cm,height=6cm,angle=270]{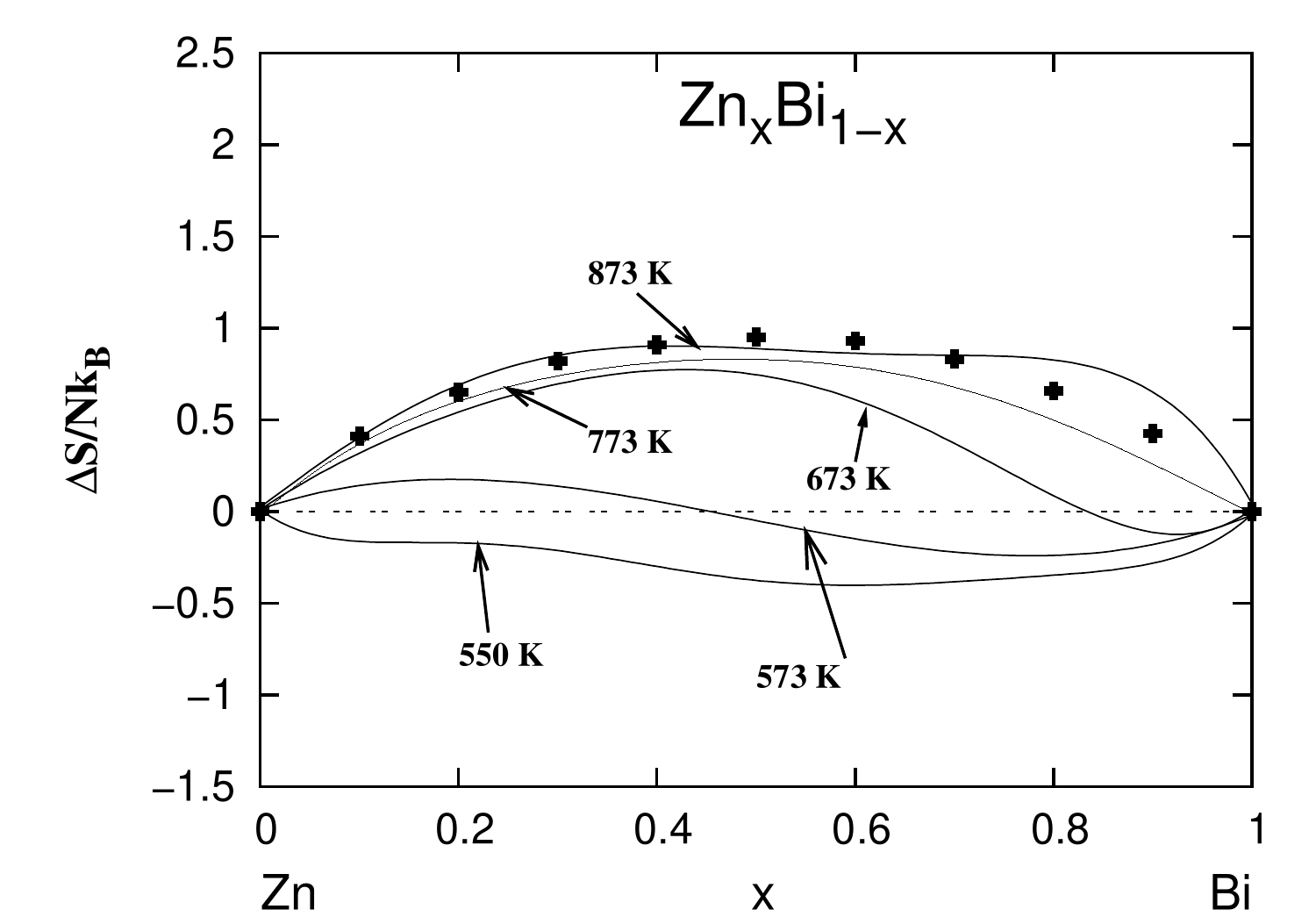}
\caption{ Entropy of mixing as a function of $x$ for Bi$_{x}$Zn$_{1-x}$ liquid binary alloys; (a) breakdown details, (b) temperature dependence (after Kasem et al. [24]).} 
\end{figure}
{(a) \it Zn$_{x}$Bi$_{1-x}$ liquid binary alloys}\\
Figure 12(a) shows the breakdown details of different contributions to the total entropy of mixing calculated by Kasem et al. [24]. It is seen from figure that at $T=$ 873 K, 
$\Delta S_{hs}$ is negative
up to $x \le 0.8$ and then becomes positive. $\Delta S_{gas}$ is negative in the concentration interval $ 0.1 < x < 0.8$, and positive beyond it. Contribution
of HSD mismatch term, $\Delta S_{\sigma}$, is almost zero across the whole range of concentration. The tail part contribution, $\Delta S_{tail}$, 
is found to be positive for the full concentration range. However, the combined effect of these contributions that is the total entropy of mixing 
is positive for all concentrations and, the agreement
 between theory and experiment is very good up to $x=0.7$, and fairly good for $x > 0.7$ [79].
\newpage
\pagestyle{fancy}
\fancyhead{}
\fancyhead[RO]{G M BHUIYAN} 
\fancyhead[LO]{\thepage}

Figure 12(b) shows the temperature dependence of entropy of mixing for Zn$_{x}$Bi$_{1-x}$ liquid binary alloys [24]. We note here that negativity of $\Delta S$
 (i.e. upward concavity) is an indication of segregation.
Figure also shows that the critical temperature and critical concentration are $T_{c}=$ 773 K and $x_{c}=0.9$, respectively. These values are found to be same as 
that found from the energy of mixing [24].\\ \\
%
\begin{figure}[!h]
\centering
\includegraphics[width=6cm,height=8cm,angle=270]{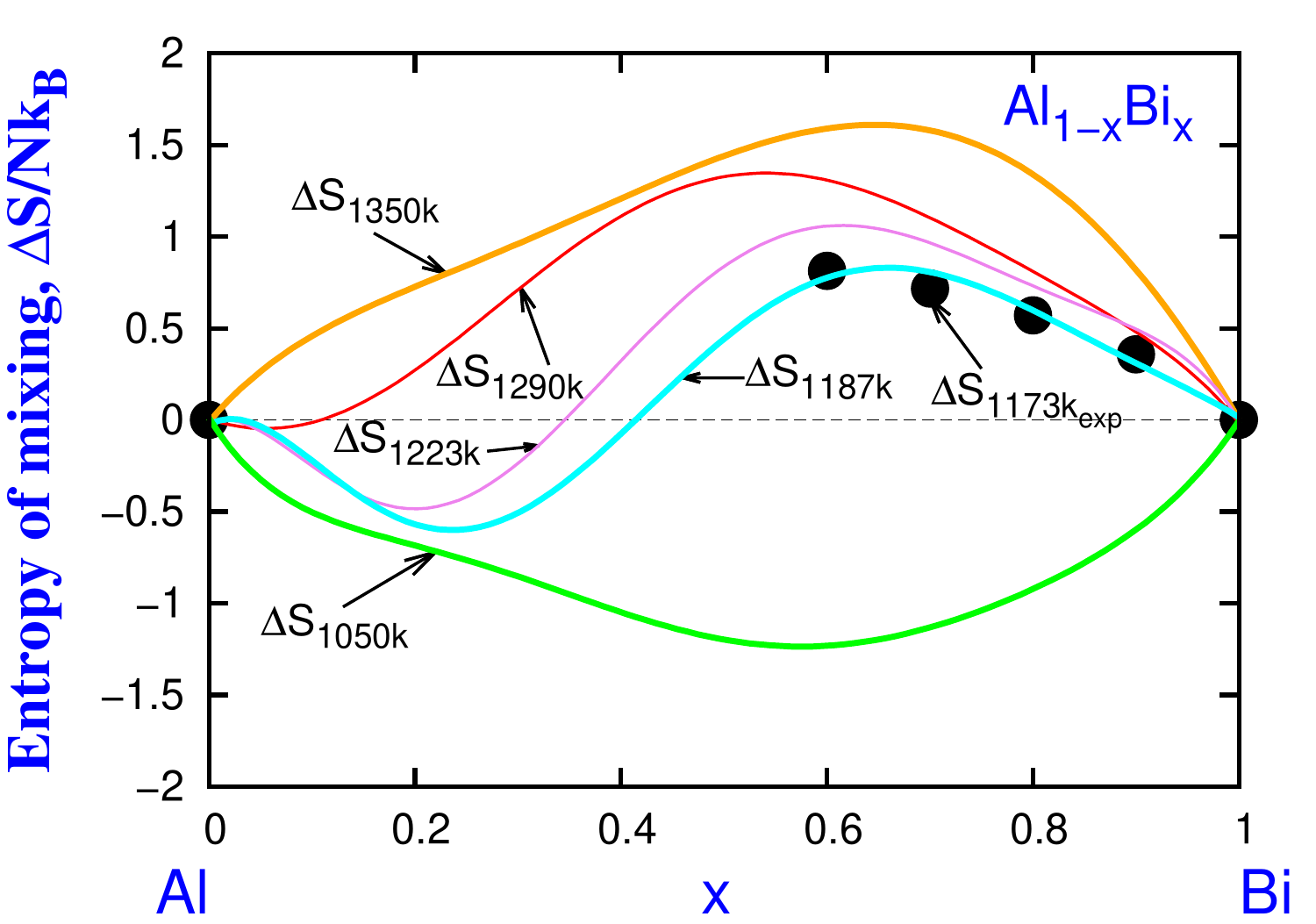}
\caption{ Temperature dependence of entropy of mixing as a function of $x$ for Bi$_{x}$Al$_{1-x}$ liquid binary alloys ( after Fysol et al. [25]).} 
\end{figure}
{(b) \it Bi$_{x}$Al$_{1-x}$ liquid binary alloys} \\
 Figure 13 shows entropy of mixing for  Bi$_{x}$Al$_{1-x}$ liquid binary alloys calculated by Fysol et al. [25]. At $T= 1350$ K, $\Delta S$ is positive for all concentrations and at $T=1050$ K 
it is negative for the whole range of concentration. For temperatures in between the entropy of mixing is partly positive and partly negative. Fysol et al. [25] theoretically 
found the values $x_{c}=$ 0.1 and $T_{c}= 1290$ K for critical concentration and critical temperature, respectively. Here the value of $T_{c}$ is the
same as that found from $\Delta F$ [25] but $x_{c}$ is somewhat lower in this case.
\section{Concluding remarks}
Looking at figures of free energy of mixing for Al$_{x}$In$_{1-x}$, Fe$_{x}$Cu$_{1-x}$, Cu$_{x}$Co$_{1-x}$, Zn$_{x}$Bi$_{1-x}$, and Bi$_{x}$Al$_{1-x}$ liquid binary alloys one can
easily find that the HS 
\newpage
\pagestyle{fancy}
\fancyhead{}
\fancyhead[LO]{MICROSCOPIC ORIGIN OF IMMISCIBILITY $\cdots$} 
\fancyhead[RO]{\thepage}
$\hspace{-6mm}$contribution $\Delta F_{hs}$ is always negative for all concentrations and temperatures. This means that, HS liquid alone cannot
describe segregation for binary alloys. This finding agrees with that of Libowitz and Rowlinson [40]. However, $\Delta F_{vol}$ becomes positive for the whole range of
 concentration and dominates other contributions in the case of Fe$_{x}$Cu$_{1-x}$, Cu$_{x}$Co$_{1-x}$, and Al$_{x}$In$_{1-x}$ liquid binary alloys; this feature directly favours  the
\begin{table}[!h]
\caption{Critical temperature and critical concentrations for different segregating alloys.}
\vspace{0.5cm}
\begin{center}
\begin{tabular}{|lcc|ccc|} 
\hline 
        &      x$_{c}$ &     &  &T$_{c}$ (K)  &    \\
Systems & (Theo.) & (Expt.)  & (Theo.) & (Expt.) & Others(Theo.) \\
\hline\\
AlIn & 0.5 & 0.5, 0.34  & 1160 & 1155, 1150, 1145 & -\\
FeCu & 0.5& 0.56, 0.538 & 1750 & 1696,1704,1694 & - \\
CuCo & 0.58 & 0.53 & 1650 & 15473& - \\
ZnBi & 0.9 & 0.83 & 773 & 856, 878 & 438 \\
BiAl & 0.15 & 0.19 & 1290 &  1310 & - \\
 \hline\\
\end{tabular}
\end{center}
\end{table}
segregation for these alloys. The contribution of the tail part of the pair potential, $\Delta F_{tail}$, becomes positive for the full concentration range for Zn$_{x}$Bi$_{1-x}$, and Bi$_{x}$Al$_{1-x}$
liquid binary alloys and negative for others. The electron gas contribution $\Delta F_{eg}$ is nearly zero for all systems and for any thermodynamic state characterized by temperature.
Energy of mixing for hard sphere liquid and the tail part contribution are very sensitive to temperature unlike $\Delta F_{vol}$ and $\Delta F_{eg}$. In the case of free energy 
$\Delta F_{hs}$ and $\Delta F_{tail}$  increases with increasing temperature, as a result total energy of mixing becomes concave downward which manifests immiscibility of the alloy.
The values of the critical temperatures and critical concentrations for different alloys are illustrated in Table 3.

Understanding of the segregating behaviour of liquid binary alloys from the microscopic theory for transport properties has just began. Some
interesting features exhibited by the coefficient of viscosity and diffusion coefficient as a function of concentration appears to be spectacular [25]. 
One of the features is the sharp bending in the $\eta$ vs $x$ (or $D$ vs $x$ ) profile around the  
\newpage
\pagestyle{fancy}
\fancyhead{}
\fancyhead[RO]{G M BHUIYAN} 
\fancyhead[LO]{\thepage}
$\hspace{-6mm}$critical concentration. Competition between the 
thermal excitation of ions and the variation of density with temperature is another one. In this case, for $T < T_{c}$, effects of atomic excitation 
dominates in determining the dynamics whereas this situation reverses for $T > T_{c}$, that is the effect of density variation with temperatures dominates the excitation effects.
The third interesting feature is the existence of a kind of scaling behaviour $(\eta_{c}- \eta) \longrightarrow (T_{c}-T)^{\beta}$ with $\beta=1.08$,
 near the critical temperature. These novel features showed by some segregating liquid binary alloys demands further research to understand the dynamic effects
in segregating alloys.
\newpage
\pagestyle{fancy}
\fancyhead{}
\fancyhead[LO]{MICROSCOPIC ORIGIN OF IMMISCIBILITY $\cdots$} 
\fancyhead[RO]{\thepage}
\section{Appendix}
The softness parameter $a_{i}$ used in the calculation are determined by fitting experimental $S(q)$ at small $q$ (see Fig.12).
\begin{figure}[!h]
\begin{minipage}{6cm}
\centering
\includegraphics[width=6cm,height=5cm]{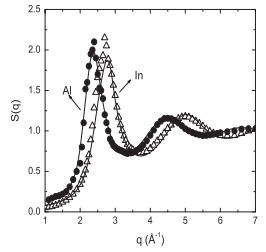}
\end{minipage}
\begin{minipage}{6cm}
\centering
\includegraphics[width=5cm,height=6cm,angle=270]{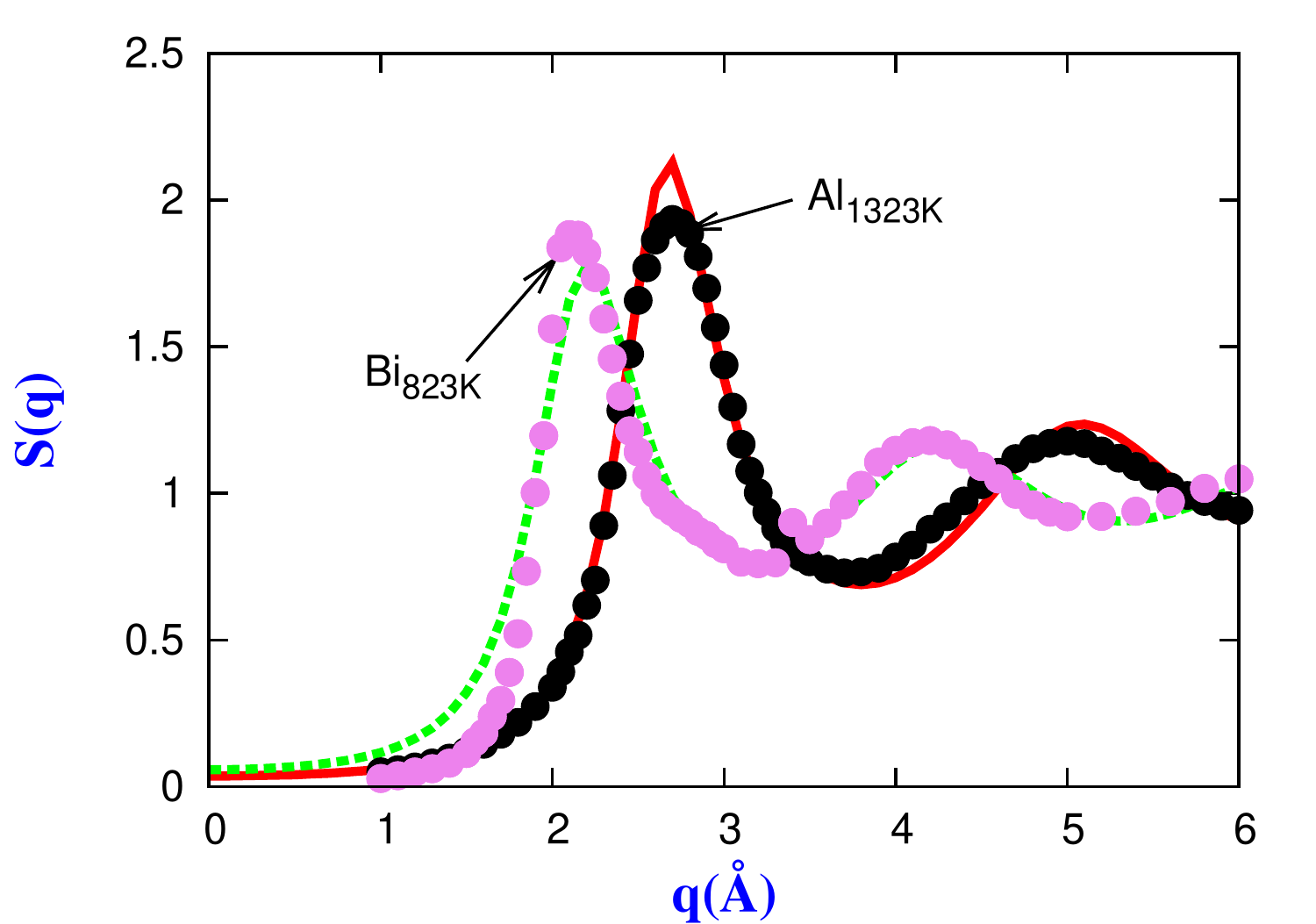}
\end{minipage}
\caption{Determination of $a_{i}$ from the best fit of S(q); line theory, closed dots experiment.$^{64)}$  For AlIn (left)(after Faruk and Bhuiyan [21] and for BiZn (right) (after Kasem et al. [24]).} 
\end{figure}
\\ \\
{\large References} \\ \\
$[1]$ B. Predel, Z. Metall {\bf 56}, 791 (1965).\\
$[2]$ H. J. Axon, Nature {\bf 162}, 997 (1948).\\
$[3]$ A. R. Miedema, P F de Chatel and F R de Boer, Physica 100B, 1 (1980).\\
$[4]$ J M Cowley, Phys. Rev. 77, 667 (1950).\\
$[5]$ B E Warren, X-ray Diffraction (Reading, M. A: Addison-Wesley, 1969). \\
$[6]$ E A Guggenheim, Mixtures (Oxford, Oxford University Press, 1952). \\
$[7]$ I Pregogine, The molecular theory of solutions (New York: North-Holland, 1957). \\
$[8]$ A. B. Bhatia and R. N. Singh, Phys. Chem. Liq. {\bf 13}, 177 (1984).\\
$[9]$ A N Campbell and R Wagemann, Can. J. Chem. 44, 657 (1966).\\
$[10]$ B Predel and H Sandig, Mat. Sci. Eng. 4, 49 (1969).\\
$[11]$ L. Ratke and S. Diefenbach, Mat. Sci and Eng. R15,263 (1995). \\
$[12]$ C D Cao, G P Gorler, D M Herlach and B Wei, Mater. Sci. Eng. A 325, 503 (2002).\\
$[13]$ D Nakagawa, Acta Metall. 6, 704 (1958).\\
$[14]$ G Wilde, R Willnecker, R N Singh and F Sommer, Z. Metall. 88, 804 
\newpage
\pagestyle{fancy}
\fancyhead{}
\fancyhead[RO]{G M BHUIYAN} 
\fancyhead[LO]{\thepage}
$\hspace{-6mm}$(1997). \\
$[15]$ C P Wang, X J Liu, I Ohnuma, R Kainuma and K Ishida, J. Phys. Edu. Diffus. 25, 320 (2004).\\
$[16]$ D V Malakhov, CALPHAD 24, 1, 2000.\\
$[17]$ G D Wignall and P A Egelstaff, J. Phys. C: Solid State Phys. 1, 1088 (1968).\\
$[18]$ Z Fan, S Ji, and J Zhang, Mater. Sci. Tech. 17, 837, 2001.\\
$[19]$ W Hoyer, I Kaban and M Merkwitz, J. optoelec. Adv. Material 5,1069 (2003).\\
$[20]$ I G Kaban, and W Hoyer, Phys. Rev. B 77,12546 (2008). \\
$[21]$ Mir Mehedi Faruk, G.M. Bhuiyan, Physica {\bf B} {\bf 422}, 156 (2013). \\
$[22]$ Mir Mehedi Faruk, G.M. Bhuiyan, Amitabh Bishwas, and Md. Sazzad Hossain. J. Chem. Phys. {\bf 140}, 134505 (2014).\\
$[23]$ D. Stroud,  Phys. Rev. {\bf B}, {\bf 7} 4405 (1973). \\
$[24]$ Md. Riad kasem, G. M. Bhuiyan and Md. Helal Uddin Maruf, J. Chem. Phys. {\bf 143}, 034503 (2015).\\
$[25]$ Fysol Ibna Abbas, G M Bhuiyan and Riad Kasem, J Phys. Soc. Japan 89, 114004 (2020).\\
$[26]$  R. N. Singh and F Sommer, Rep. Prog. Phys. {\bf 60}, 57 (1997).\\
$[27]$  J Blanco, D J Gonzalez, L E Gonzalez, J M Lopez and M J Stott, Phys Rev. E 67, 041204 (2003). \\
$[28]$ H Rupersberg and W Knoll, Z Naturfor a 32, 1314 (1977).\\
$[29]$ N March, S Wilkins and J E Tibbals, Cryst. Latt. Defects 6, 253 (1976). \\
$[30]$ A B Bhatia and R N Singh, Phys. Chem. Liq. 11, 285 (1982).\\
$[31]$ N H March, Phys. Chem. Liq. 20, 241 (1989). \\
$[32]$ Y A Odusote, L A Hussain, O E Awe, J Non-Cryst. Solids 353, 1167 (2007).\\
$[33]$ M. Shimoji,  {\it Liquid Metals:  An Introduction to the Physics and Chemistry of Metals in the liquid State} ( Academic Press, London. 1977). \\
$[34]$ G M Bhuiyan and A Z Ziauddin Ahmed,, Physica B 390, 377 (2007).\\
$[35]$ G. M. Bhuiyan, Md. Saiful Alam, A. Z. Ziauddin  Ahmed, Istiaque M. Syed and R. I. M. A. Rashid, J. Chem. Phys. {\bf 131} (2009).  \\             
$[36]$ D. Stroud, Phys. Rev. {\bf B}, {\bf 8}, 1308 (1973).\\
$[37]$ J.Hafner, {\it From Hamiltonoians to Phase Diagrams} (Springer, Berlin, 1987) p. {\bf 56}.\\
$[38]$ I. H. Umar, A. Meyer, M. Watabe and W. H. Young, J. Phys. F. {\bf 4}, 1691 (1974a).\\
\newpage
\pagestyle{fancy}
\fancyhead{}
\fancyhead[LO]{MICROSCOPIC ORIGIN OF IMMISCIBILITY $\cdots$} 
\fancyhead[RO]{\thepage}
$\hspace{-6mm}$$[39]$ J. L. Lebowitz, Phys. Rev. {\bf 133}, A895 (1964).\\
$[40]$ J. L. Lebowitz and J.S. Rowlison, J. Chem. Phys. 41, 133 (1964).\\
$[41]$ S. A. Rice and A. R. Alnatt, J. Chem. Physics. {\bf 34}, 2144 (1961).\\
$[42]$ A. R. Alnatt and S. A. Rice, J. Chem. Physics. {\bf 34}, 2156 (1961).\\
$[43]$ M. Kitajima, T. Itami and M. Shimoji, Phil. Mag. {\bf 30},285 (1974).\\
$[44]$ N W Ashcroft and N D Mermin, Solid State Physics (Holt, Rinehart and Winston, 1976)\\
$[45]$  M P Marder, Condensed Matter Physics (John Wiley and Sons, Hoboken, Newjersey, 2010) \\
$[46]$ R M Zief, G E Uhlenbeck and M Kac, Phys. Report 32, 169 (1977).\\
$[47]$ M. W. Zymansky and R. H. Dittman, {\it Heat and Thermodynamics} (The Mcgraw-Hill Company Inc., New York, 1997) 7th ed..  \\ 
$[48]$ R K Pathria, Statistical Mechanics (Pergamon Press, Oxford, 1985).\\
$[49]$ Mehran Kardar, Statistical Physics for Particles (Cambridge University Press, Chembridge, 2007). \\
$[50]$  N. W. Ashcroft and D. C. Lagrength, Phys. Rev. {\bf 156}, 685 (1967). \\
$[51]$ J. D. Weeks, D. Chandler and H.C. Andersen, J. Chem. Phys. {\bf 55}, 5422-5423 (1971). \\
$[52]$ A. Meyer, M. Silbert, and W. H. Young, Chem. Phys. {\bf 49},147 (1984).\\
$[53]$ H C Andersen, D Chandler and J D Weeks, Adv. Chem. Phys. 34, 105 (1976).\\
$[54]$ G D Mahan, Many-particle Physics, (Plenum Press, New York, 1983) p 459.\\
$[55]$ P Protopapas. H C Andersen and N A D Parlee, J. Chem. Physics 59, 15 (1973).\\
$[56]$ J. L. Bretonnet and M. Silbert, Phys. Chem. Liq. {\bf 24}, 169 (1992).\\
$[57]$ A B Bhatia and D E Thornton, Phys. Rev. B 9, 435 (1970).\\
$[58]$ R N Singh and F Sommer, Z. Metall. 83, 7 (1992).\\
$[59]$ S. Ichimaru, K. Utsumi, Phys. Rev. B {\bf 24}, 7385 (1981).\\
$[60]$ G. M. Bhuiyan, J. L. Bretonnet, L.E.  Gonz\'alez, M. Silbert, J. Phys. Condens. Matter {\bf 4}, {\bf 7651} (1992).\\          
$[61]$ G. M. Bhuiyan, J. L. Bretonnet and M. Silbert, J. Non-Cryst. Solids 145, 156 (1993).\\
$[62]$ M. A. Khaleque, G. M. Bhuiyan, S. Sharmin, R. I. M. A. Rashid, S. M. Mujibur Rahman, Eur.Phys.J. {\bf B} {\bf 26}, 319 (2002). \\
$[63]$ J. L. Bretonnet, G. M. Bhuiyan and M. Silbert, J. Phys. Condens. Matter {\bf 4}, 5359 (1992).\\
\newpage
\pagestyle{fancy}
\fancyhead{}
\fancyhead[RO]{G M BHUIYAN} 
\fancyhead[LO]{\thepage}
$\hspace{-6mm}$$[64]$ Fysol Ibna Abbas, G. M. Bhuiyan and Md. Riad kasem, J. Non-Cryst. Solids {\bf 481}, 391 (2018). \\
$[65]$ F. Zahid, G. M. Bhuiyan, S. Sultana, M.A. Khaleque, R.I.M.A. Rashid and S. M. Mujibur Rahman, Phys. Status Sol. B {\bf 215}, 987 (1999).\\
$[66]$ G. M. Bhuiyan, I. Ali and S. M. Mujibur Rahman, Physics B {\bf 334},147 (2003).\\
$[67]$ E. H. Bhuiyan, A. Z. Ziauddin Ahmed, G. M. Bhuiyan M. Shahjahan, Physica B {\bf 403},1695 (2008).  \\
$[68]$ S Chanda, A Z Ziauddin Ahmed, G M Bhuiyan, S K Barman and S Sarker, J. Non-Cryst. Solids 357, 3774 (2011).\\
$[69]$ A B Patel and H Sheng, Phys. Rev. B 102, 064101 (2020).\\
$[70]$ M A Mohaiminul Islam, R C Gosh and G M Bhuiyan, J. Mol. Liq. 290, 111224 (2019).\\
$[71]$ S. Sharmin, G. M. Bhuiyan, M. A. Khaleque, R.I.M.A. Rashid and S. M. Mujibur Rahman, Phys. Status Sol B {\bf 232},243 (2002).\\
$[72]$ M D Salah Uddin, R C Gosh and G M Bhuiyan, J Non-Cryst. Solids 499, 426 (2018).\\
$[73]$ D. A. McQuarrie, {\it Statistical Mechanics} (Harper and Row, New york, 1976).\\
$[74]$ S. Tamaki, Phys. Lett. A {\bf 40}, 17 (1972).\\
$[75]$ N W Ashcroft, Phys. Lett. A 23, 48 (1966).\\
$[76]$ J A Moriarty, Phys. Rev. B 16, 2537 (1977).\\
$[77]$ J Hubbard, Proc. R. Soc. Lond. 243, 336 (1957).\\
$[78]$ Y. Waseda, {\it The Structure of Non-Crystalline Materials} (McGraw-hill, 1984).\\
$[79]$ R. Hultgren, P. D. Desai, D. T. Hawking, M. Gleiser and K. K. Keluey, {\it Selected values of Thermodynamic Properties of Binary Alloys} (American Society of Metals, 1973).\\
$[80]$ A N Campbell, L B Buchamann, J M Kuzmzk, R H Tuxworth, J. Am. Chem. Soc. 74, 1609 (1952).\\
$[81]$ B Predel, Z. Metallk, 56, 791 (1965).\\
$[82]$ F Sommer, H G Krull and S K Yu in: L Ratke (Ed), Immiscible Liquid Metals and Organics, DGM- Informationsgesellschaft, (1993) p.79.\\
$[83]$ A. J. McAllster,  Bull. Alloy Phase Diag. {\bf 5}, 247 (1984).  \\
$[84]$ V. Raghavan, J. Phase Equil. Diffus. {\bf 33}, 166 (2012).\\
$[85]$ S Karlhuber, A Mikula, R N Singh and F Sommer, J Alloys and Comp. 283, 198 (1999).\\


\end{document}